\documentclass[a4paper,11pt]{article}
\pdfoutput=1

\usepackage{jinstpub}
\usepackage{amsmath,amssymb}

\title{Background rejection in NEXT using deep neural networks}

\author[a,1]{J.~Renner,\note{Corresponding author.}}
\author[b]{A.~Farbin,}
\author[a]{J.~Mu\~{n}oz Vidal,}
\author[a]{J.M.~Benlloch-Rodr\'{i}guez,}
\author[a]{A.~Botas,}
\author[a]{P.~Ferrario,}
\author[a,2]{J.J.~G\'omez-Cadenas,\note{Co-spokesperson.}}
\author[a]{V.~\'{A}lvarez,}
\author[h]{C.D.R.~Azevedo,}
\author[c]{F.I.G.~Borges,}
\author[a]{S.~C\'{a}rcel,}
\author[a]{J.V.~Carri\'{o}n,}
\author[d]{S.~Cebri\'{a}n,}
\author[a]{A.~Cervera,}
\author[c]{C.A.N.~Conde,}
\author[a]{J.~D\'{i}az,}
\author[p]{M.~Diesburg,}
\author[f]{R.~Esteve,}
\author[c]{L.M.P.~Fernandes,}
\author[h]{A.L.~Ferreira,}
\author[c]{E.D.C.~Freitas,}
\author[e]{A.~Goldschmidt,}
\author[q]{D.~Gonz\'{a}lez-D\'{i}az,}
\author[i]{R.M.~Guti\'{e}rrez,}
\author[j]{J.~Hauptman,}
\author[c]{C.A.O.~Henriques,}
\author[k]{J.A.~Hernando Morata,}
\author[f]{V.~Herrero,}
\author[b]{B.~Jones,}
\author[l]{L.~Labarga,}
\author[a]{A.~Laing,}
\author[p]{P.~Lebrun,}
\author[a]{I.~Liubarsky,}
\author[a]{N.~L\'{o}pez-March,}
\author[a]{D.~Lorca,}
\author[i]{M.~Losada,}
\author[r]{J.~Mart\'{i}n-Albo,}
\author[k]{G.~Mart\'{i}nez-Lema,}
\author[a]{A.~Mart\'{i}nez,}
\author[a]{F.~Monrabal,}
\author[c]{C.M.B.~Monteiro,}
\author[f]{F.J.~Mora,}
\author[h]{L.M.~Moutinho,}
\author[a]{M.~Nebot-Guinot,}
\author[a]{P.~Novella,}
\author[b,2]{D.~Nygren,}
\author[a]{B.~Palmeiro,}
\author[p]{A.~Para,}
\author[l]{J.~P\'{e}rez,}
\author[a]{M.~Querol,}
\author[n]{L.~Ripoll,}
\author[a]{J.~Rodr\'{i}guez,}
\author[c]{F.P.~Santos,}
\author[c]{J.M.F.~dos Santos,}
\author[a]{L.~Serra,}
\author[e]{D.~Shuman,}
\author[a]{A.~Sim\'{o}n,}
\author[o]{C.~Sofka,}
\author[a]{M.~Sorel,}
\author[f]{J.F.~Toledo,}
\author[n]{J.~Torrent,}
\author[g]{Z.~Tsamalaidze,}
\author[h]{J.F.C.A.~Veloso,}
\author[o,3]{J.~White,\note{Deceased.}}
\author[o]{R.~Webb,}
\author[a]{N.~Yahlali,}
\author[i]{H.~Yepes-Ram\'{i}rez}

\affiliation[a]{Instituto de F\'{i}sica Corpuscular (IFIC), CSIC \& Universitat de Val\`{e}ncia,\\ 
	Calle Catedr\'{a}tico Jos\'{e} Beltr\'{a}n, 2, Paterna, Valencia, 46980 Spain}
\affiliation[b]{University of Texas at Arlington\\
	701 S. Nedderman Drive, Arlington, Texas, 76019 USA}
\affiliation[c]{Departamento de Fisica, Universidade de Coimbra, \\
	Rua Larga, Coimbra, 3004-516 Portugal}
\affiliation[d]{Lab. de F\'{i}sica Nuclear y Astropart\'{i}culas, Universidad de Zaragoza,\\
	Calle Pedro Cerbuna, 12, Zaragoza, 50009 Spain}
\affiliation[e]{Lawrence Berkeley National Laboratory (LBNL),\\
	1 Cyclotron Road, Berkeley, California, 94720 USA}
\affiliation[f]{Instituto de Instrumentaci\'{o}n para Imagen Molecular (I3M), Universitat Polit\`{e}cnica de Val\`{e}ncia,\\
	Camino de Vera, s/n, Edificio 8B, Valencia, 46022 Spain}
\affiliation[g]{Joint Institute for Nuclear Research (JINR),\\
	Joliot-Curie, 6, Dubna, 141980 Russia}
\affiliation[h]{Institute of Nanostructures, Nanomodelling and Nanofabrication (i3N), Universidade de Aveiro,\\
	Campus de Santiago, Aveiro, 3810-193 Portugal}
\affiliation[i]{Centro de Investigaciones, Universidad Antonio Nari\~{n}o,\\
	Carretera 3 este No. 47A-15, Bogot\'{a}, Colombia}
\affiliation[j]{Department of Physics and Astronomy, Iowa State University,\\
	12 Physics Hall, Ames, Iowa, 50011-3160 U.S.A.}
\affiliation[k]{Instituto Gallego de F\'{i}sica de Altas Energ\'{i}as (IGFAE), Univ. de Santiago de Compostela,\\
	Campus sur, R\'{u}a Xos\'{e} Mar\'{i}a Su\'{a}rez N\'{u}\~{n}ez, s/n, Santiago de Compostela, 15782 Spain}
\affiliation[l]{Departamento de F\'{i}sica Te\'{o}rica, Universidad Aut\'{o}noma de Madrid,\\
	Ciudad Universitaria de Cantoblanco, Madrid, 28049 Spain}
\affiliation[m]{Dpto. de Mec\'{a}nica de Medios Continuos y Teor\'{i}a de Estructuras, Univ. Polit\`{e}cnica de Val\`{e}ncia,\\
	Camino de Vera, s/n, Valencia, 46071 Spain}
\affiliation[n]{Escola Polit\`{e}cnica Superior, Universitat de Girona,\\
	Av. Montilivi, s/n, Girona, 17071 Spain}
\affiliation[o]{Department of Physics and Astronomy, Texas A\&M University,\\
	College Station, Texas, 77843-4242 U.S.A.}
\affiliation[p]{Fermi National Accelerator Laboratory,\\
	Batavia, Illinois, 60510 U.S.A.}
\affiliation[q]{CERN, European Organization for Nuclear Research,\\
	Geneva, 1211 Switzerland}
\affiliation[r]{Department of Physics, University of Oxford,\\
	Denys Wilkinson Building, Keble Road, Oxford OX1 3RH, United Kingdom}

\emailAdd{jrenner@ific.uv.es}

\newcommand{\bb}{\ensuremath{\beta\beta}}
\newcommand{\bbonu}{\ensuremath{0\nu\beta\beta}}
\newcommand{\bbtnu}{\ensuremath{2\nu\beta\beta}}

\newcommand{\mbb}{\ensuremath{m_{\bb}}}
\newcommand{\Tonu}{\ensuremath{T_{1/2}^{0\nu}}}

\newcommand{\Qbb}{\ensuremath{Q_{\bb}}}

\newcommand{\ckky}{\ensuremath{\mathrm{cts~keV^{-1}~kg^{-1}~yr^{-1}}}}

\newcommand{\XE}{\ensuremath{^{136}\mathrm{Xe}}}

\newcommand{\TL}{\ensuremath{^{208}\mathrm{Tl}}}
\newcommand{\BI}{\ensuremath{^{214}\mathrm{Bi}}}

\newcommand{\THT}{\ensuremath{^{232}\mathrm{Th}}}
\newcommand{\UTT}{\ensuremath{^{238}\mathrm{U}}}

\newcommand{\CHF}{\ensuremath{\mathrm{CH}_4}}
\newcommand{\CFF}{\ensuremath{\mathrm{CF}_4}}

\abstract{We investigate the potential of using deep learning techniques to reject background 
events in searches for neutrinoless double beta decay with high pressure xenon time projection 
chambers capable of detailed track reconstruction.  The differences in the topological signatures of background and signal events can be learned by deep neural networks via training over many thousands of events.  These networks can then be used to classify further events as signal or background, providing an additional background rejection factor at an acceptable loss of
efficiency.  The networks trained in this study performed better than previous methods developed based on the use of the
same topological signatures by a factor of 1.2 to 1.6, and there is potential for further improvement.}


\arxivnumber{1609.06202}

\collaboration{\includegraphics[height=9mm]{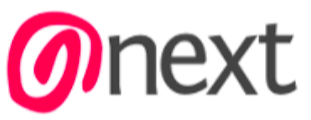}\\[6pt]
   NEXT collaboration}

\begin{document} 
\maketitle
\flushbottom

\section{The NEXT experiment}\label{sec:intro}
Double beta decay with neutrino emission (\bbtnu) is a  process in which two simultaneous $\beta$ decays occur within a nucleus,

\begin{equation}
 (Z,A) \rightarrow (Z+2,A) + 2e^{-} + 2\bar{\nu}_{e}.
\end{equation}

This process is allowed in the Standard Model and has been observed in several isotopes.  Double-beta decay has also been
postulated to exist in the zero-neutrino mode, or neutrinoless double beta decay (\bbonu), in which the two 
antineutrinos are not emitted and the total energy released in the decay, \Qbb, is 
carried away by the two electrons.  The observation of \bbonu\ would imply that the neutrino is its own anti-particle, that is, a
Majorana particle \cite{Schechter_1982}, amongst other important physical implications (see for example \cite{GomezCadenas:2013ue, Cadenas_2012, Avignone_2008}).

After 75 years of experimental effort, no compelling evidence for the existence of \bbonu\ decay has been obtained.  For a given isotope, the lifetime of \bbonu\,\,decay depends on a nuclear 
matrix element and a phase-space integral, both of 
which can be calculated to some uncertainty, and the square of the effective neutrino mass $|m_{\beta\beta}|^2 = |\sum_{i=e,\nu,\tau}U_{ei}^2m_{i}|^2$ which is a combination of the neutrino
masses $m_{i}$ and neutrino mixing matrix elements $U_{ei}$.  The lifetime is of the order of $10^{25}-10^{26}$~years for a degenerate neutrino mass hierarchy ($m_1 \sim m_2 \sim m_3$), $10^{26}-10^{27}$~years for an inverted 
neutrino mass hierarchy ($m_3 \ll m_1 < m_2$), and longer than $10^{27}$~years  for a normal mass hierarchy ($m_1 < m_2 \ll m_3$).  Experiments of the current generation deploy approximately
100 kg of the candidate isotope and are subject to several tens of counts per year of background events in their region of interest (ROI) of energy selection near $Q_{\beta\beta}$ \cite{GomezCadenas:2013ue}.  These experiments
will be capable of probing only the parameter space corresponding to the degenerate mass hierarchy, perhaps pushing into the inverted hierarchy.  The most sensitive lower bound to date was set by the KamLAND-Zen experiment with $^{136}$Xe, at $T_{1/2}^{0\nu} > 1.06 \times 10^{26}$ years \cite{KamLANDZen_2016}. In order to completely cover the parameter 
space of the inverted mass hierarchy, experiments employing candidate isotope masses at the tonne-scale with background rates of (at most) a few counts per tonne-year will be required  \cite{Gomez-Cadenas:2015twa}.  

One of the technologies currently being developed is that of high pressure xenon (HPXe) Time Projection Chambers (TPCs). In particular, the NEXT collaboration is building a HPXe TPC capable of 
containing a total mass of 100 kg of xenon enriched at 90\% in the \bb\ decaying isotope \XE \cite{NEXT_sensitivity}. This detector, called NEXT-100, will operate at 15 bar and use electroluminescent 
(EL) amplification of the ionization signal to optimize energy resolution. The detection of EL light provides an energy measurement using 60 photomultipliers (PMTs) located behind the cathode 
(the \emph{energy plane}) as well as 
tracking via a dense array of about 8,000 silicon photomultipliers (SiPMs) located behind the anode (the \emph{tracking plane}).
In addition to performing a competitive search for \bbonu, NEXT-100 will explore potential techniques for operation and background rejection at the tonne-scale.  

The NEXT background model predicts a background rate of $4 \times 10^{-4}$~\ckky\ in the ROI  \cite{NEXT_sensitivity}. The energy resolution for NEXT-100 is assumed to be 0.7\% FWHM ($\sim$ 17 keV) at \Qbb. The experiment expects, therefore, less than one count of background per 100 kg and year of exposure, and thus its sensitivity to \Tonu\ is not dominated by background subtraction and increases rapidly with exposure. The expected sensitivity to the \bbonu\ half-life is $\Tonu > 6 \times 10^{25}$~yr for an exposure of 275 kg$\cdot$yr. This translates into a \mbb\ sensitivity range of $[90-180]$~meV, depending on the nuclear matrix element.

The NEXT collaboration has already built and tested several kg-scale prototypes, NEXT-DBDM \cite{Alvarez:2012kua} and
NEXT-DEMO \cite{Alvarez:2012xda,Alvarez:2013gxa,Lorca:2014sra}, which have both demonstrated the excellent energy resolution (extrapolated to 0.5--0.7\% FWHM at
$Q_{\beta\beta}$) obtainable in high pressure xenon gas.  NEXT-DEMO has demonstrated the feasibility of signal/background discrimination based on the topology of reconstructed tracks \cite{NEXT_topology},
an essential component to identifying \bbonu\,events and rejecting background events (see section \ref{sec.topology}).  The collaboration is currently commissioning the first underground phase of the experiment, the so called NEXT-W{\sc{hite}} (or NEW for short\footnote{The name honours the memory of the late Professor James White, whose knowledge and generosity were essential to launching the experiment.}). NEW deploys a  mass of 10 kg of xenon at 15 bar, the energy plane hosts 12 PMTs and the tracking plane nearly 2,000 SiPMs. Operation is foreseen in 2016 and 2017, while NEXT-100 is scheduled to start operations in 2018.

A central feature of a HPXe TPC is the capability of imaging electron tracks providing a {\bf topological signature} that can be used to separate signal events (the two electrons emitted in a \bbonu\ decay) from background events (mainly due to single electrons with kinetic energy comparable to the end-point of the \bbonu\ decay, \Qbb). In this paper, we study the performance of the topological signature, analyzing how it is affected by the various physics processes involved in the propagation of electrons in dense gas, as well as by the detector spatial resolution. We use both the conventional reconstruction of electrons in NEXT described in \cite{NEXT_topology}, and an alternative technique based on the use of deep neural networks (DNNs), comparing their performance. 

\section{Imaging tracks in a HPXe-EL TPC}
\label{sec.topology}

\begin{figure}[!htb]
\centering
\includegraphics[width= 0.95\textwidth]{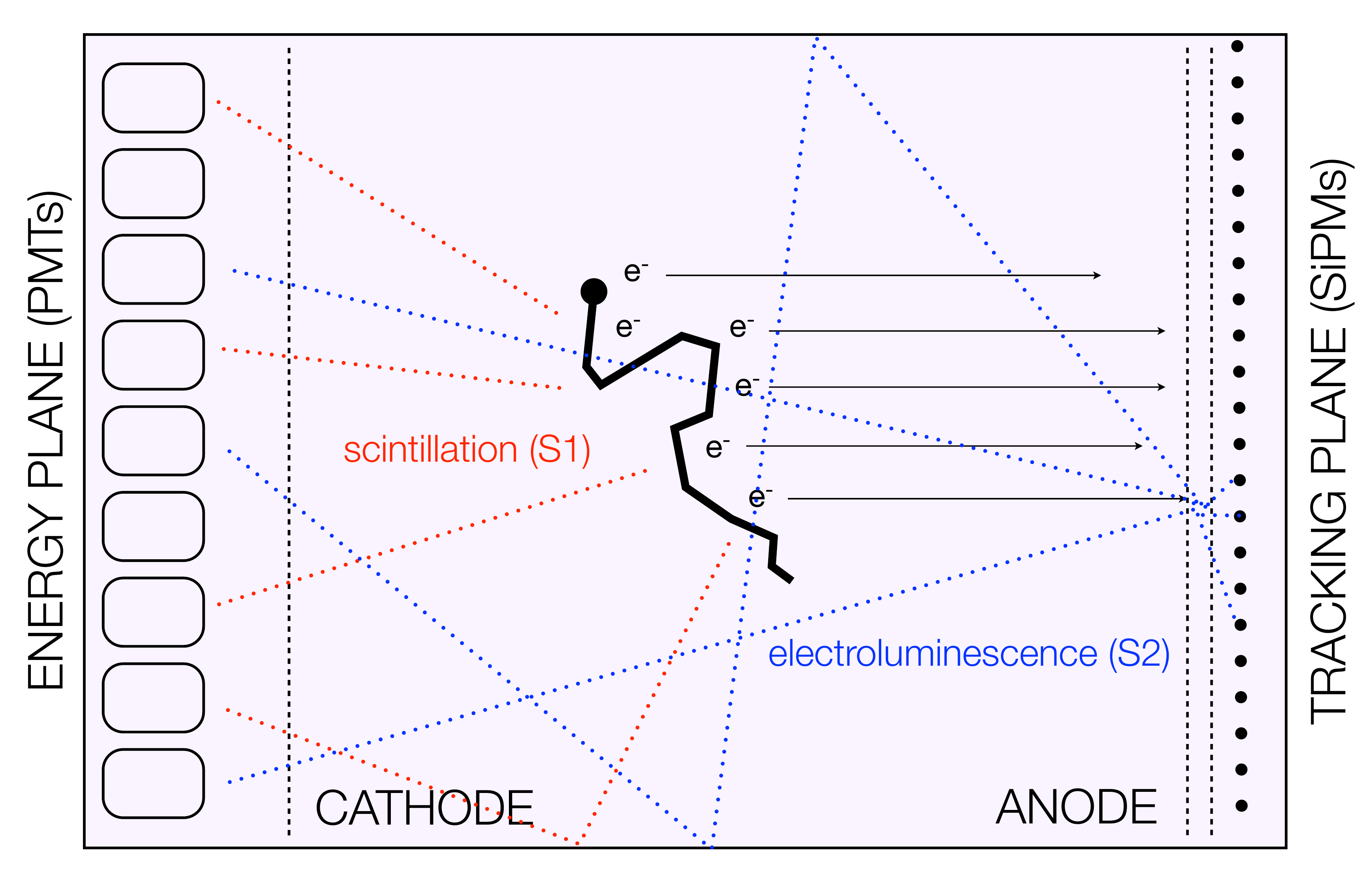}
\caption{Principle of operation of an asymmetric HPXe TPC with EL readout.  (Figure from \cite{Alvarez:2013gxa}.)} \label{fig.SS}
\end{figure}

Figure \ref{fig.SS} shows the principle of operation of an asymmetric HPXe TPC using proportional electroluminescent (EL) amplification of the ionization signal (as is the case for NEXT-100). The detection process involves the use of the prompt scintillation light ($S_1$) from the gas as the start-of-event time, and the drift of the ionization charge to the anode by means of an electric field ($\sim0.3$ kV/cm at 15 bar) where secondary EL scintillation ($S_2$) is produced in a narrow region defined by a highly transparent mesh and a quartz plate coated with ITO (indium tin oxide) and TPB (tetraphenyl butadiene), called the EL gap.  High voltages are applied to the two meshes to establish an electric field of $\sim 20$ kV/cm at 15 bar in this region. The detection of EL light provides an energy measurement using PMTs in the case of NEXT-100 located behind the cathode (the \emph{energy plane}). The reconstruction of the track topology is carried out with a dense array of SiPMs located behind the anode (the \emph{tracking plane}). The $x$-$y$~coordinates are found using the information provided by the tracking plane, while $z$~is determined by the drift time between the detection of $S_1$~and $S_2$. For each reconstructed spatial point, the detector also measures the energy deposited. Thus, the track is imaged as a collection of hits, and each hit is defined by a 3D space coordinate and by an associated energy deposition, as $(x, y, z, E)$.

\begin{figure}[!htb]
\centering
\includegraphics[width= 0.95\textwidth]{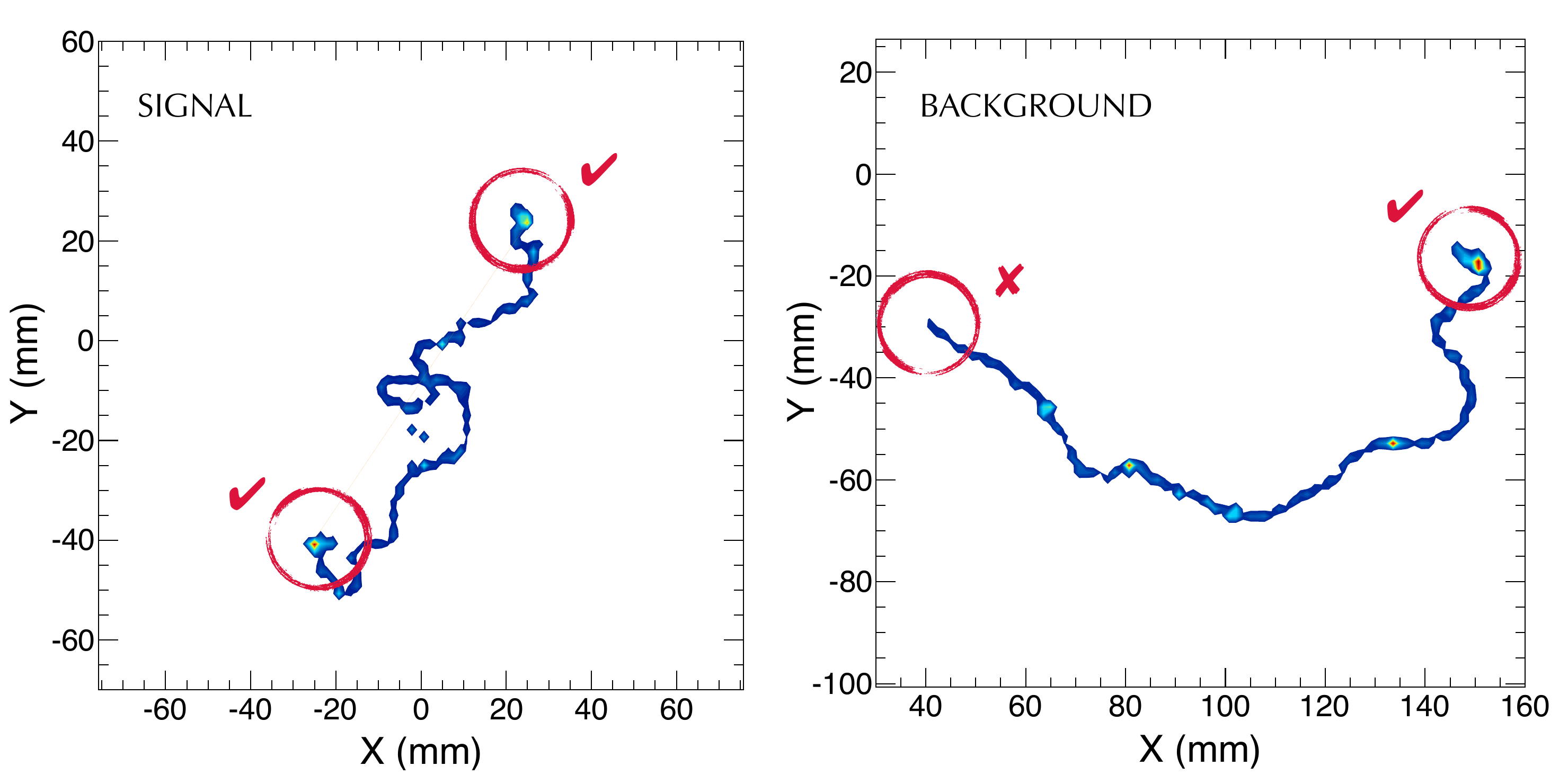}
\caption{Monte Carlo simulation of a signal (\bbonu) event (left) and a background event (right) in xenon gas at 15~bar. The color corresponds to energy deposition in the gas, red representing higher density of energy deposition and blue representing lower density. The signal consist of two electrons emitted from a common vertex, and thus it features large energy depositions (blobs) at both ends of the track. Background events are, typically, single-electron tracks (produced by photoelectric or Compton interactions of high energy gammas emitted by \BI\ or \TL\ isotopes), and thus feature only one blob (figure from \cite{NEXT_sensitivity}).} \label{fig.ETRK2}
\end{figure}

\begin{figure}[!htb]
\centering
\includegraphics[width=0.45\textwidth]{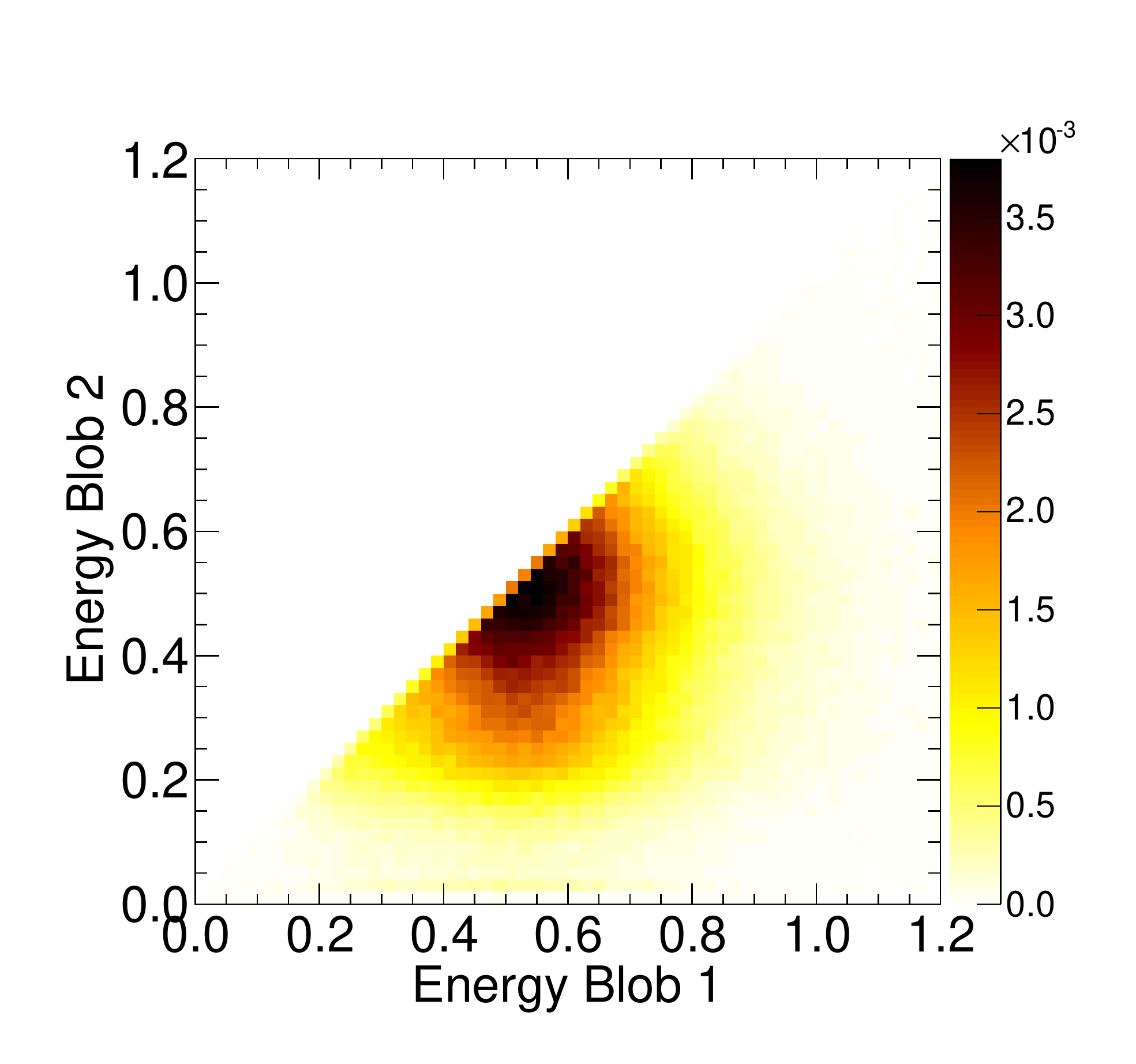}
\includegraphics[width=0.45\textwidth]{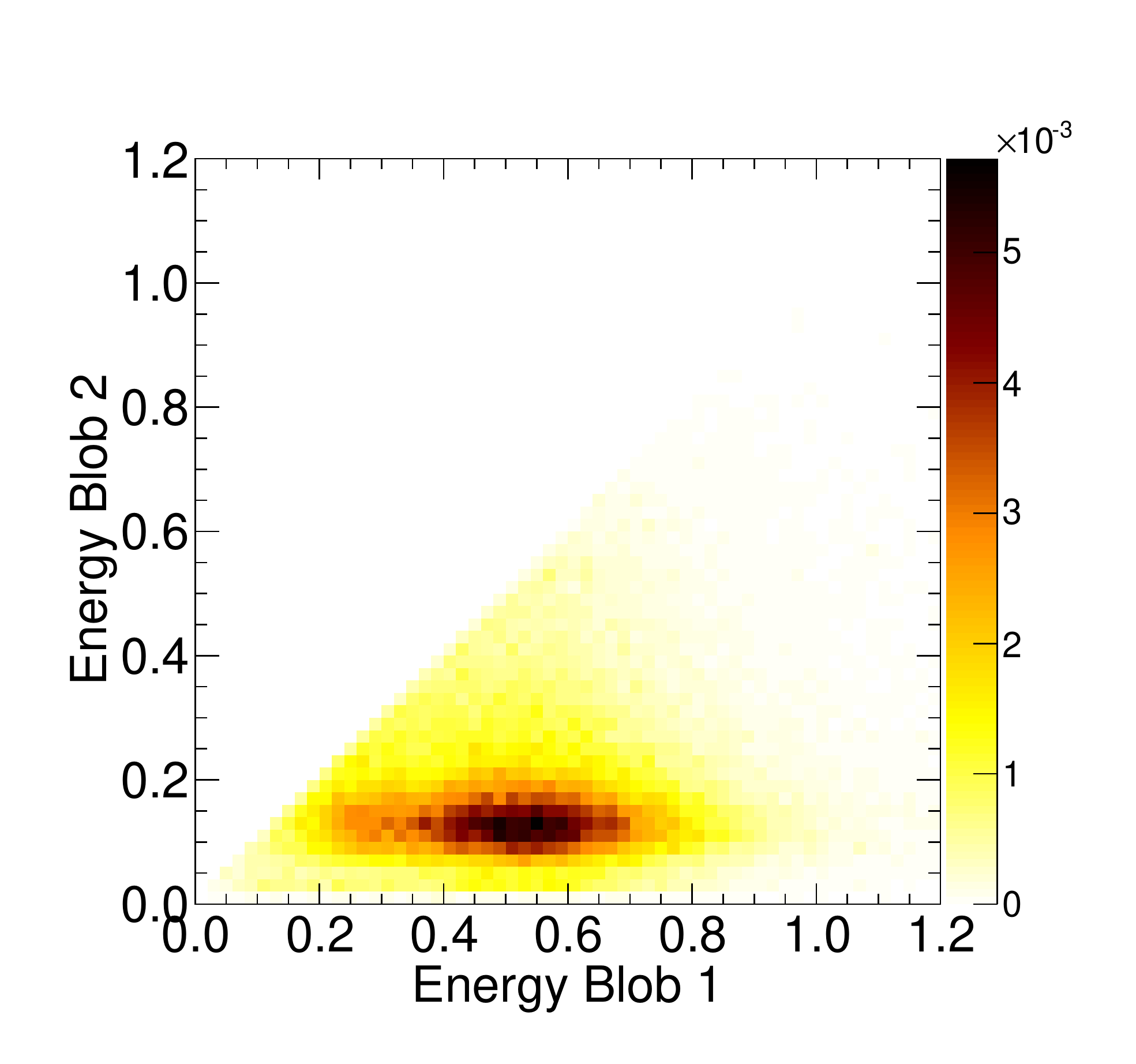}
\caption{Probability distribution of signal (left) and background (right) events in terms of the energies of the end-of-track blob candidates. The blob candidate labelled as `1' corresponds to the more energetic one, whereas `blob 2' corresponds to the less energetic of the two. In a signal event, the blob candidates have, on average, the same energy. In a background event, blob candidate 1 has an energy similar to that of a signal event while the energy of blob candidate 2 is very small (figure from \cite{NEXT_sensitivity}).} \label{fig.BLOBS}
\end{figure}

Electrons (and positrons) moving through xenon gas lose energy at an approximately fixed rate until they become non-relativistic. At the end of the trajectory the $1/v^2$~ rise of the energy loss (where $v$~ is the speed of the particle) leads to a significant energy deposition in a compact region, which will be referred to as a ``blob''. The two electrons produced in double beta decay events appear as a single continuous trajectory with a blob at each end (figure \ref{fig.ETRK2}, left). The main background in NEXT comes from high energy gammas emitted in \TL\ and \BI\ decays, which occur naturally in the detector materials as part of the \THT\ and \UTT\ chains and enter the active volume of the detector. These gammas convert in the gas through photoelectric, Compton and pair production processes. Except in the case of pair production, these electrons typically leave a single continuous track with only one blob (figure \ref{fig.ETRK2}-right). This topological signature was used in the Gotthard TPC to give an overall rejection rate of 96.5\% for single-electron events in high-pressure xenon gas at 5 atm pressure \cite{Gotthard_1998}. Likewise, in NEXT, reconstruction of the signal and background topology using the tracking plane provides a powerful means of background rejection. For each track, the energy in the regions at both extremes of the track is measured and labelled as $E_{b,1}$ (the energy of the most energetic blob candidate), and $E_{b,2}$ (the energy of the least energetic blob candidate). In a signal event, $E_{b,1} \sim E_{b,2} $, while for background events $E_{b,1} >> E_{b,2} $. Figure \ref{fig.BLOBS} shows how this feature can be used to separate signal from background.

\subsection{Reconstruction of tracks in a HPXe-EL TPC}\label{ssec.reconstruction}
Reconstruction of tracks in an electroluminescent HPXe TPC is complicated by the diffusion of the charge cloud during drift and also by the nature of the read-out. Scintillation light is produced over the whole width of the EL gap (5 mm in NEXT-100) spreading the signal from a single electron over a time inversely proportional to the drift velocity within the gap ($\sim$ 2 $\mu$s). Additionally, the EL light is produced isotropically and, therefore, the signal produced by the passage of an electron through the gap is expected to arrive at the tracking plane ($\sim$ several mm behind the anode) over the area defined by the intersection of the plane with the sphere of light.

In a previous paper \cite{Lorca:2014sra}, the NEXT collaboration demonstrated that a ``point-like''
deposition of charge due to the absorption of a point-like source (such as the xenon K$_\alpha$ X-rays) can be parameterised as a two dimensional Gaussian with a standard deviation of $\sim$ 8 mm where the spread due to EL light production is the dominant effect with subdominant contributions from transverse diffusion of the charge. As discussed below, the resolution with which we can reconstruct the centroid of this optical spread function is significantly better than this. Longitudinally, the expected spread has a noticeable dependence on the drift distance since the diffusion dominates. K$_\alpha$ events are expected to have widths in $z$ with standard deviations of between 0.5 mm, for very short drifts, to about 5 mm for the longest drifts. In order to optimise the reconstruction of tracks these values must be taken into account by dividing the signal information into appropriate time slices and using charge information from clustered SiPM channels.

The standard NEXT algorithm searches for clusters around local maxima and then proceeds iteratively, selecting first the channel with maximum charge and forming a cluster with the first ring of sensors around it. The cluster information is then used to build a hit, whose $x$ and $y$ position are reconstructed as the barycentre of the charge information. 

Once a set of hits is found, a connectivity criterium is  defined so that the hits belonging to each separate particle can be grouped into tracks. The procedure is as follows: first, the active volume is divided into 3D pixels, known as ``voxels'', of fixed dimensions. Each voxel is given an energy equal to the sum of the energies of all the hits which fall within its boundaries. The collection of voxels obtained in such a way can be regarded as a graph, defined as a set of nodes and links that connect pairs of nodes. Two voxels can then be considered connected if they share a face, an edge or a corner, with each pair of connected voxels being given a weight equal to the geometric distance between their centres. Next, the ``Breadth First Search'' (BFS) algorithm (see for example \cite{Cormen_2001}) is used to group the voxels into tracks and to find their end-points and length. The BFS algorithm is a graph search algorithm which finds the minimum path between two connected nodes, starting from one node and exploring all its neighbours first, then the second level neighbours and so on, until it reaches the second node. The BFS algorithm divides the voxels into connected sets, known as tracks, and finds their end-points, defined as the pair of voxels with largest distance between them, where the distance of two voxels is the shortest path that connects them. The distance between the end-points is the length of the track. See \cite{NEXT_topology} for a thorough discussion.

The choice of the voxel size is a compromise between a fine granularity and conservation of connectivity, which depends on the hit-finder algorithm in use. In \cite{NEXT_topology},  the best connectivity  was  found for voxels of 
$10 \times 10 \times 10 \mathrm{~mm^3}$. The analysis described in \cite{NEXT_sensitivity} used voxels
of similar size ($10 \times 10 \times 5 \mathrm{~mm^3}$).  Improvements in the hit-finder algorithm (or the use of alternative methods such as DNNs) may allow, in principle, for smaller voxels. However, the size of the voxel also reflects the effect of the spatial resolution, which in turn depends on: 

\begin{enumerate}
\item {\em Tracking plane segmentation:} this includes the pitch of the SiPMs in the tracking plane as well as the SiPM response. Indeed, the use of SiPMs with very low dark current and high gain allows one to determine the location of an event by weighting the position of each sensor with the light recorded, thus improving dramatically the ``digital'' resolution, which goes as $\sim$pitch/$\sqrt{12}$. The digital point resolution corresponding to a pitch of 10 mm (NEXT-100) is 10$/\sqrt{12} \sim 3$~mm. Using weighted information (e.g, local barycenter algorithms), the point resolution improves to about 1 mm (see \cite{Lorca:2014sra}). 
\item {\em The width of the EL region:}
The non-zero width of the EL gap adds an extra resolution term in the $z$~coordinate which goes like $w/\sqrt{12}$~where $w$~is the width of the grid. For the NEXT-100 detector, $w\sim 5$ mm, resulting also in a resolution of about 1.5 mm. At the same time, the non-negligible distance (several mm) between the gate grid (the plane that defines the beginning of the EL region) and the sensors in the tracking plane, spreads the signal of a single electron over several SiPMs. The light distribution is Gaussian and a fit to the profile recovers the position of the ionization electron.
\item {\em Diffusion of the drifting electron cloud:} both transverse and longitudinal diffusion are high in pure xenon (of the order of 10 and 5 mm/$\sqrt{\rm{m}}$, respectively). On the other hand, work in progress within the NEXT collaboration \cite{Azevedo:2015eok} suggests that adding small amounts of cooling gases such as \CHF\ or \CFF\ to pure xenon (at the level of 0.1 \% of \CHF\ or 0.01\% of \CFF) reduces both transverse and longitudinal diffusion to some $2.0$~mm/$\sqrt{\rm{m}}$. This is one of the most important upgrades under study for the second phase of NEXT-100. 
\end{enumerate}

\section{Monte Carlo Simulation}\label{sec.MC}
NEXUS \cite{MartinAlbo_thesis}, the  Geant4-based \cite{GEANT4} Monte Carlo simulation of the NEXT experiment, permits an accurate modelling of the detector geometry and provides the tools to carry out both full and fast simulations of the apparatus response. A fast simulation has been chosen for this study, given the need to generate a very large number of events for the detailed physics studies presented here, as well as for the training of the DNNs.

The simulation begins by generating a large number of signal and background events. Neutrinoless double beta events, 2 electrons with momenta generated according to a distribution calculated by the DECAY0 code \cite{Ponkratenko_2000} for $0\nu\beta\beta$ decay, are randomly created throughout the active region of the detector, while the leading background events --- gamma rays of energies 2.447 MeV and 2.614 MeV corresponding to gammas emitted by daughters of $^{214}$Bi and $^{208}$Tl, respectively (see \cite{NEXT_sensitivity} for a thorough discussion) --- are shot from the field cage of the detector geometry.  The resulting locations and magnitudes of the energy depositions in the active volume are recorded as ``true hits''.  A minimum step size of 1 mm is used in NEXUS.

The fast-simulation approach to producing reconstructed objects, starting from the true hits, takes into account the energy and spatial resolution. The former is introduced by simply smearing the total energy deposited by the event by the expected NEXT-100 resolution (we assume, conservatively, 0.7\% FWHM at \Qbb); the latter, by combining the resolution associated with the pixel pitch, the EL width and the diffusion into the voxel size. Thus, to emulate the response of NEXT-100 operating with pure xenon, the true hits are replaced by voxels of $10 \times 10 \times 5 \mathrm{~mm^3}$. Comparison between fast and full simulation results in \cite{NEXT_topology}, with a similar voxel size of $10 \times 10 \times 10 \mathrm{~mm^3}$ as used here, showed that the efficiencies obtained for the classification cut (see section \ref{ssec:ClassificationAnalysis}) are in agreement within 5\%. These efficiency results are also in agreement with the measured value. For this reason, we believe that a voxelization of size $10 \times 10 \times 5 \mathrm{~mm^3}$ is a good proxy to capture all spatial resolution effects in a pure xenon detector with tens of centimeters of drift. In addition, we consider an optimistic scenario of $2 \times 2 \times 2 \mathrm{~mm^3}$ voxels. Spatial resolution studies using full simulation and reconstruction have also been performed with the baseline NEXT-100 EL region and tracking plane design, based on a 1 cm pitch between SiPMs and an EL gap of 5 mm, and single-point spatial resolutions of order 1 mm have been obtained in this case. This is comparable with the single-point resolution introduced by 3D voxels of size 2 mm, given by $2\sqrt{3} / \sqrt{12}$ mm = 1 mm.

\section{The topological signature}
\label{sec.top}

\subsection{Pre-selection of data}
\label{ssec.prep}

 Once the list of voxels is obtained for each event, the data is processed as follows: 
\begin{enumerate}
	\item[1.] Only events near \Qbb\ (in the energy window between 2.4 and 2.5 MeV) are accepted.
	\item[2.] A fiducial cut is applied, ensuring that no more than 10 keV of energy was deposited within 2 cm of the edges of the active region.
	\item[3.] Tracks are formed using the BFS algorithm and only events with exactly one track are accepted.  This cut effectively suppresses background events which interacted by Compton scattering followed by photoelectric conversion, and also those accompanied by the emission of x-rays associated with the de-excitation of the xenon atom (e.g, after a photoelectric interaction). 
\end{enumerate}

Table \ref{tbl.FastAnalysisResults} shows how the set of cuts described above reduces the signal and the primary backgrounds for Monte Carlo generated events with voxel sizes of $2 \times 2 \times 2 \mathrm{~mm^3}$ and $10 \times 10 \times 5 \mathrm{~mm^3}$ (events were generated originating from the field cage surrounding the active region of the detector).  Notice that, in order to characterize the rejection power of the topological signature, a relatively large energy window of 100 keV around \Qbb\ is used. The total rejection power of NEXT will be the combination of the rejection power achieved by the pre-selection (cuts 1-2 above), the topological cuts (cut 3 above and the classification cut, discussed in more detail below), and a final, stricter energy cut that accepts events in a relatively narrow ROI around \Qbb. See \cite{NEXT_sensitivity} for a detailed discussion.

\begin{table}[!htb]
	\begin{center}
		\caption[Fast analysis summary]{\label{tbl.FastAnalysisResults}Fraction of events remaining after each analysis cut, for signal events ($10^5$ initial events generated within the active region of the detector) and background events from $^{208}$Tl ($10^{9}$ initial events) and $^{214}$Bi ($10^{10}$ initial events) generated from the field cage surrounding the active region.}
		\begin{tabular}{c|cc|cc|cc}
			\\
			 & \multicolumn{2}{c}{\textbf{Signal Events}} & \multicolumn{2}{c}{\textbf{BG Events ($^{208}$Tl)}} & \multicolumn{2}{c}{\textbf{BG Events ($^{214}$Bi)}}\\
			\textbf{Cut} & $2 \times 2 \times 2$ & $10 \times 10 \times 5$ & $2 \times 2 \times 2$ & $10 \times 10 \times 5$ & $2 \times 2 \times 2$ & $10 \times 10 \times 5$\\
			\hline
			(Initial events) & 1.0 & 1.0 & 1.0 & 1.0 & 1.0 & 1.0\\
			Energy & $7.59$\small{$\times 10^{-1}$} & $7.59$\small{$\times 10^{-1}$} & $2.27$\small{$\times 10^{-3}$} & $2.27$\small{$\times 10^{-3}$} & $1.42$\small{$\times 10^{-4}$} & $1.42$\small{$\times 10^{-4}$}\\
			Fiducial & $6.71$\small{$\times 10^{-1}$} & $6.68$\small{$\times 10^{-1}$} & $1.19$\small{$\times 10^{-3}$} & $1.17$\small{$\times 10^{-3}$} & $8.62$\small{$\times 10^{-5}$} & $8.54$\small{$\times 10^{-5}$}\\
			Single-Track & $3.75$\small{$\times 10^{-1}$} & $4.79$\small{$\times 10^{-1}$} & $7.90$\small{$\times 10^{-6}$} & $1.81$\small{$\times 10^{-5}$} & $3.84$\small{$\times 10^{-6}$} & $8.75$\small{$\times 10^{-6}$}\\
			Classification* & $3.23$\small{$\times 10^{-1}$} & $3.67$\small{$\times 10^{-1}$} & $7.70$\small{$\times 10^{-7}$} & $2.41$\small{$\times 10^{-6}$} & $2.90$\small{$\times 10^{-7}$} & $9.59$\small{$\times 10^{-7}$}\\
		\end{tabular}
	\end{center}
	* See section \ref{ssec:ClassificationAnalysis}
\end{table}

\subsection{The standard NEXT classification analysis}
\label{ssec:ClassificationAnalysis}
After pre-selection and the initial topological cut eliminating events with multiple connected tracks, the events were classified as signal or background based on the presence of one or 
two ``blobs'' of energy in the reconstructed track.  All possible shortest paths between two voxels were found using the BFS algorithm, and the first and last voxels of the longest of such 
paths\footnote{Note that such a path 
may not have included, in fact most likely did not include, all voxels in the event.} were considered to be the beginning and end of the track.  For both the beginning and end of the track, a 
``blob'' candidate was constructed by summing the energy of all voxels located within a given ``radius'' $r_b$ of the corresponding beginning or end voxel.  Note that distances between voxels
were computed using the shortest path distance as determined by the BFS algorithm and not using Euclidean distance, so the quantity $r_b$ should not be thought of literally as the radius of
a sphere containing the ``blob'' candidate.  The use of such a summation avoids, in many cases, the duplication of voxels in the two ``blob'' candidates.  Such duplication could be present if the track wrapped around such 
that its end was located within a short Euclidean distance of its beginning.  The summations yielded two energies, $E_{b,1}$, assigned to the greater of the two energies, and $E_{b,2}$.

\begin{figure}[!htb]
	\centering
	\includegraphics[scale=0.36]{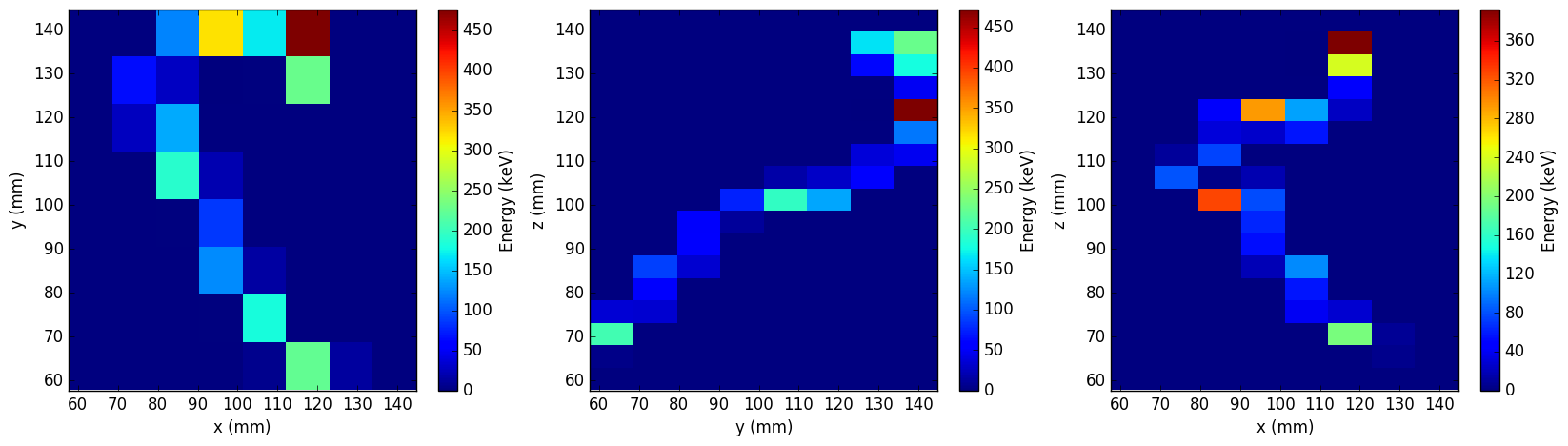}
	\includegraphics[scale=0.36]{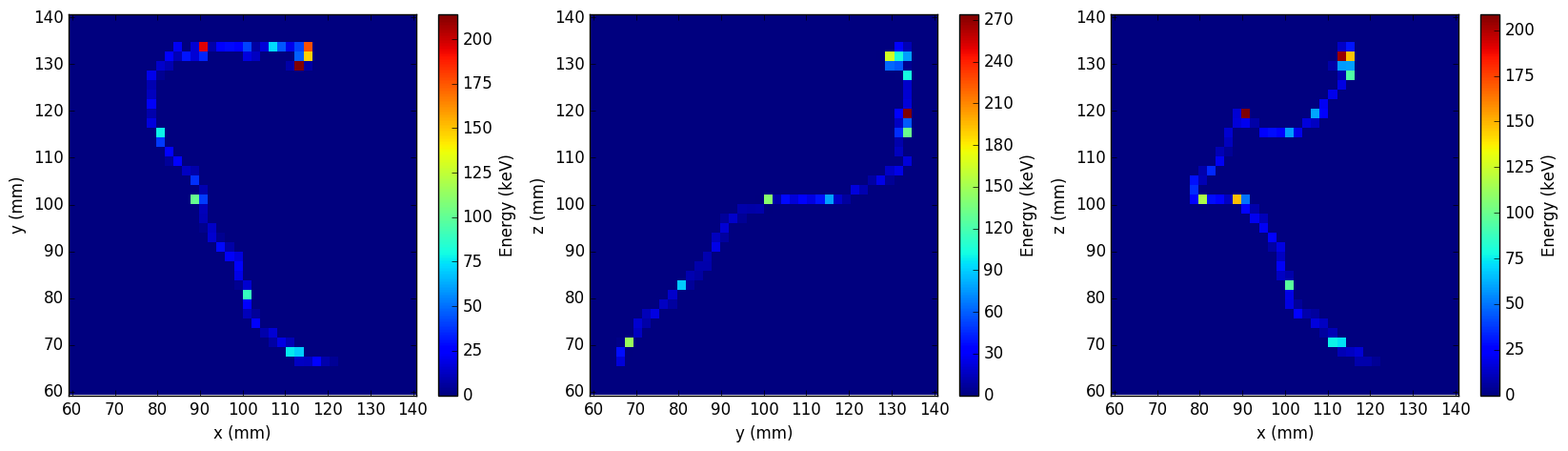}
	\caption{\label{fig.exampleProjs10105}Projections in xy, yz, and xz for an example background event voxelized with $10 \times 10 \times 5$ mm$^3$ voxels (above) and with $2 \times 2 \times 2$ mm$^3$ voxels (below).}
\end{figure}

\begin{figure}[!htb]
	\centering
	\includegraphics[scale=0.36]{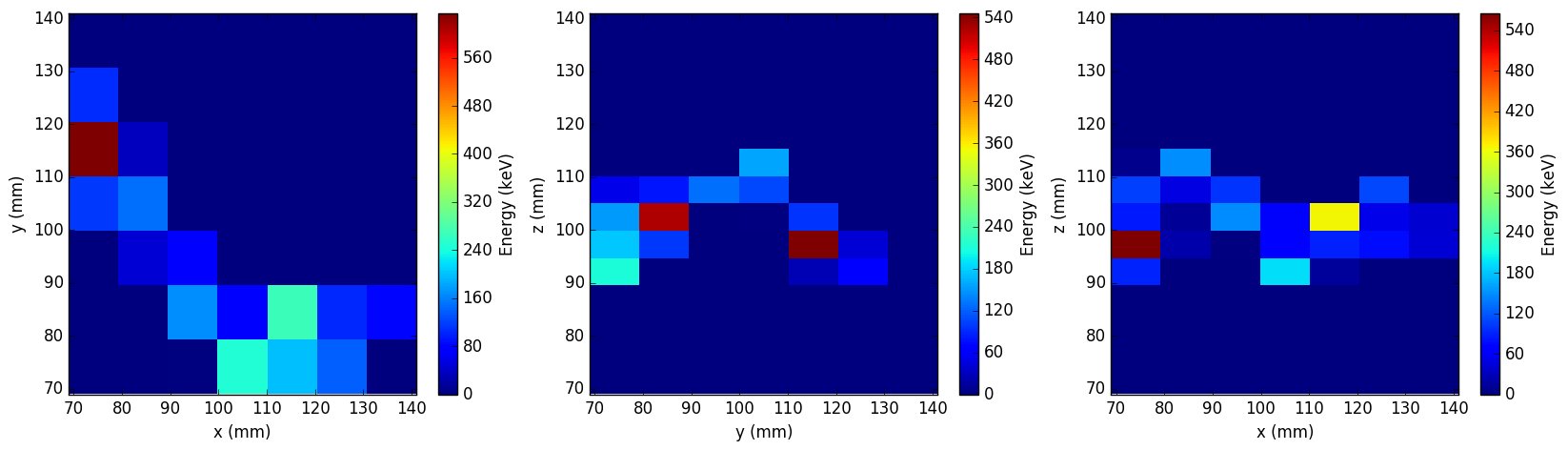}
	\includegraphics[scale=0.36]{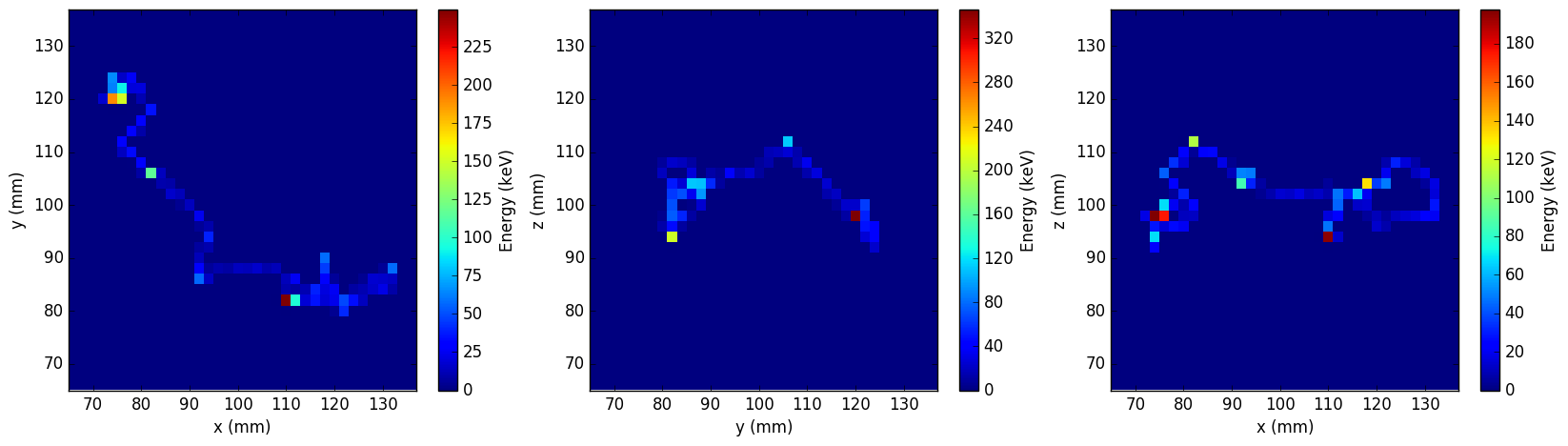}
	\caption{\label{fig.exampleProjs222}Projections in xy, yz, and xz for an example signal event voxelized with $10 \times 10 \times 5$ mm$^3$ voxels (above) and with $2 \times 2 \times 2$ mm$^3$ voxels (below).}
\end{figure}

\begin{figure}[!htb]
	\centering
	\includegraphics[scale = 0.35]{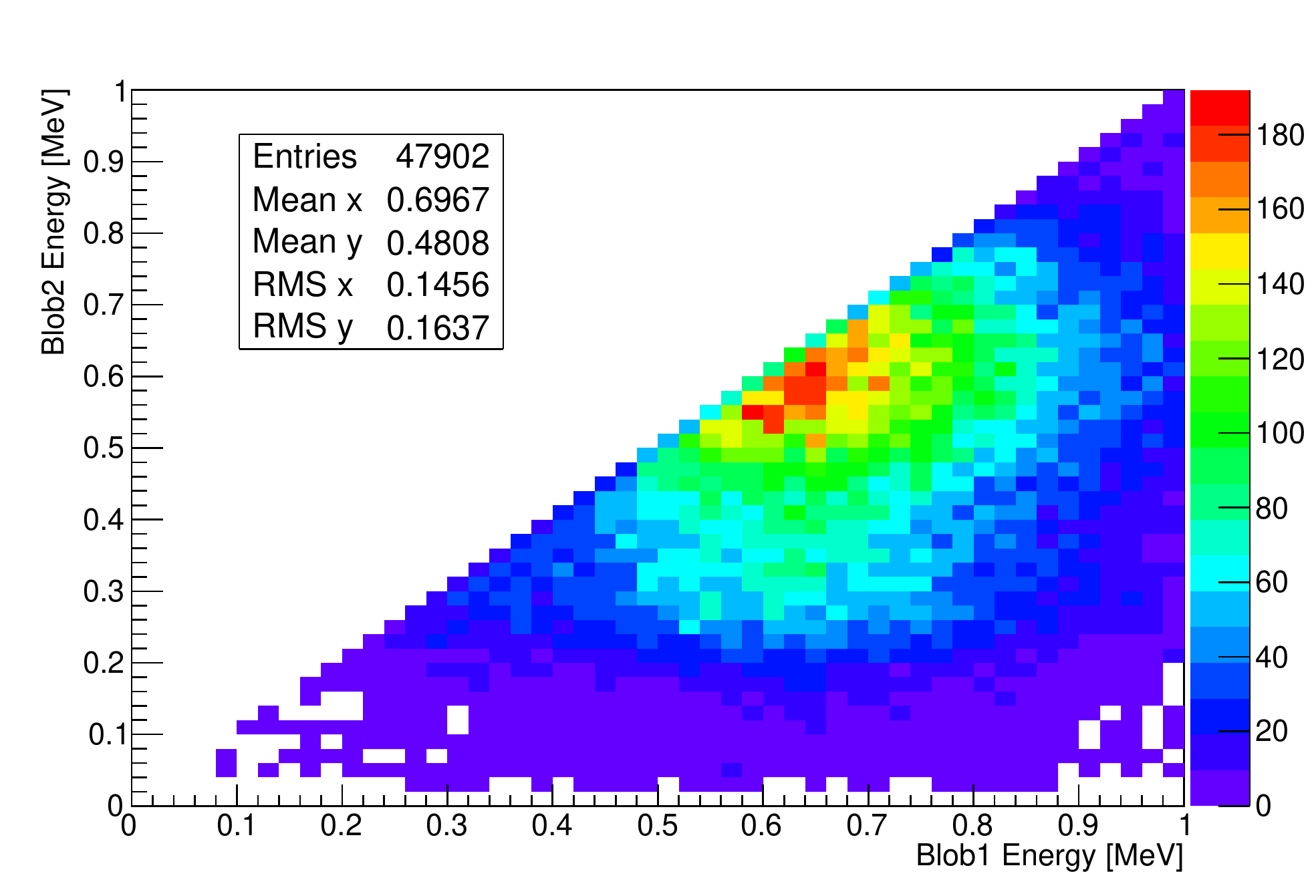}
	\includegraphics[scale = 0.35]{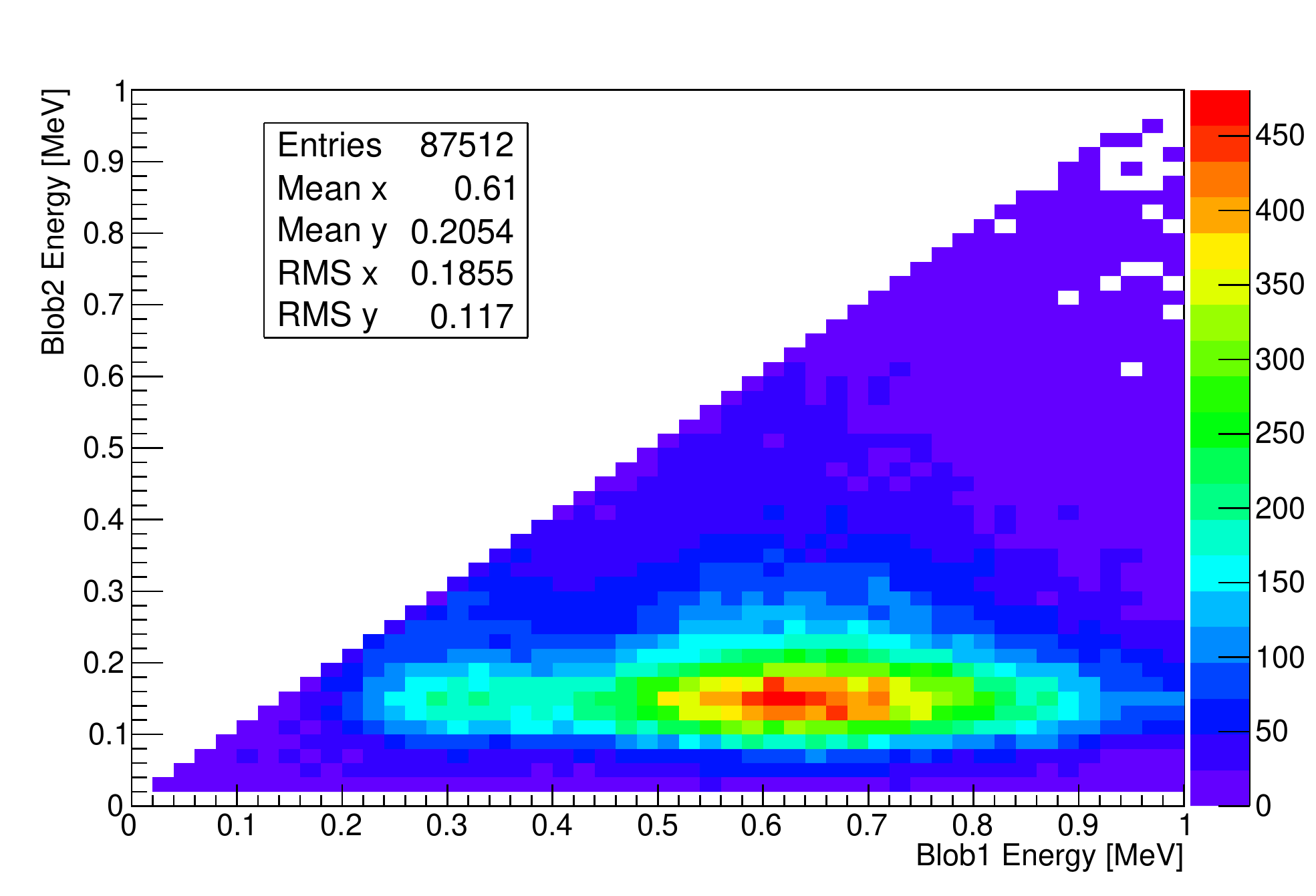}	
	\includegraphics[scale = 0.35]{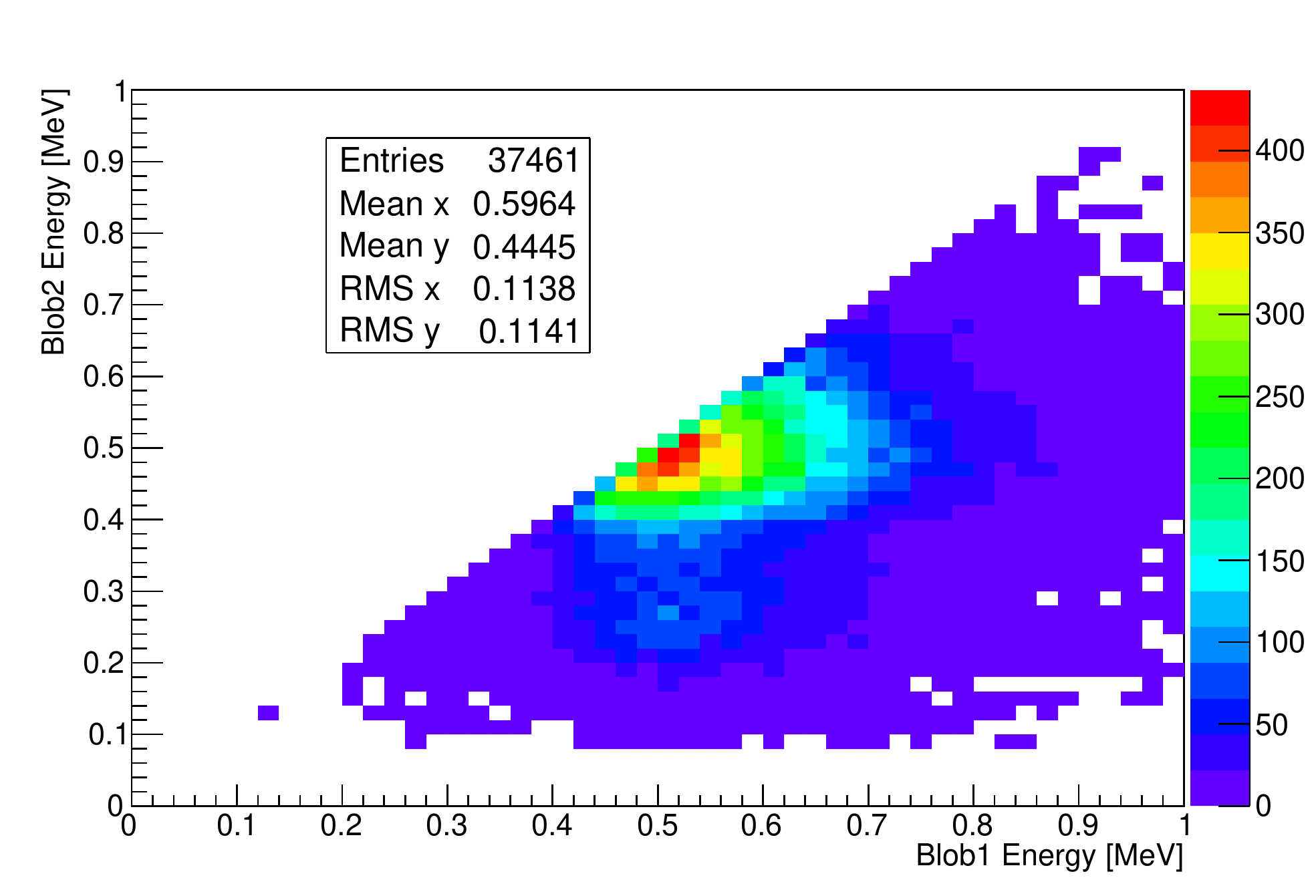}
	\includegraphics[scale = 0.35]{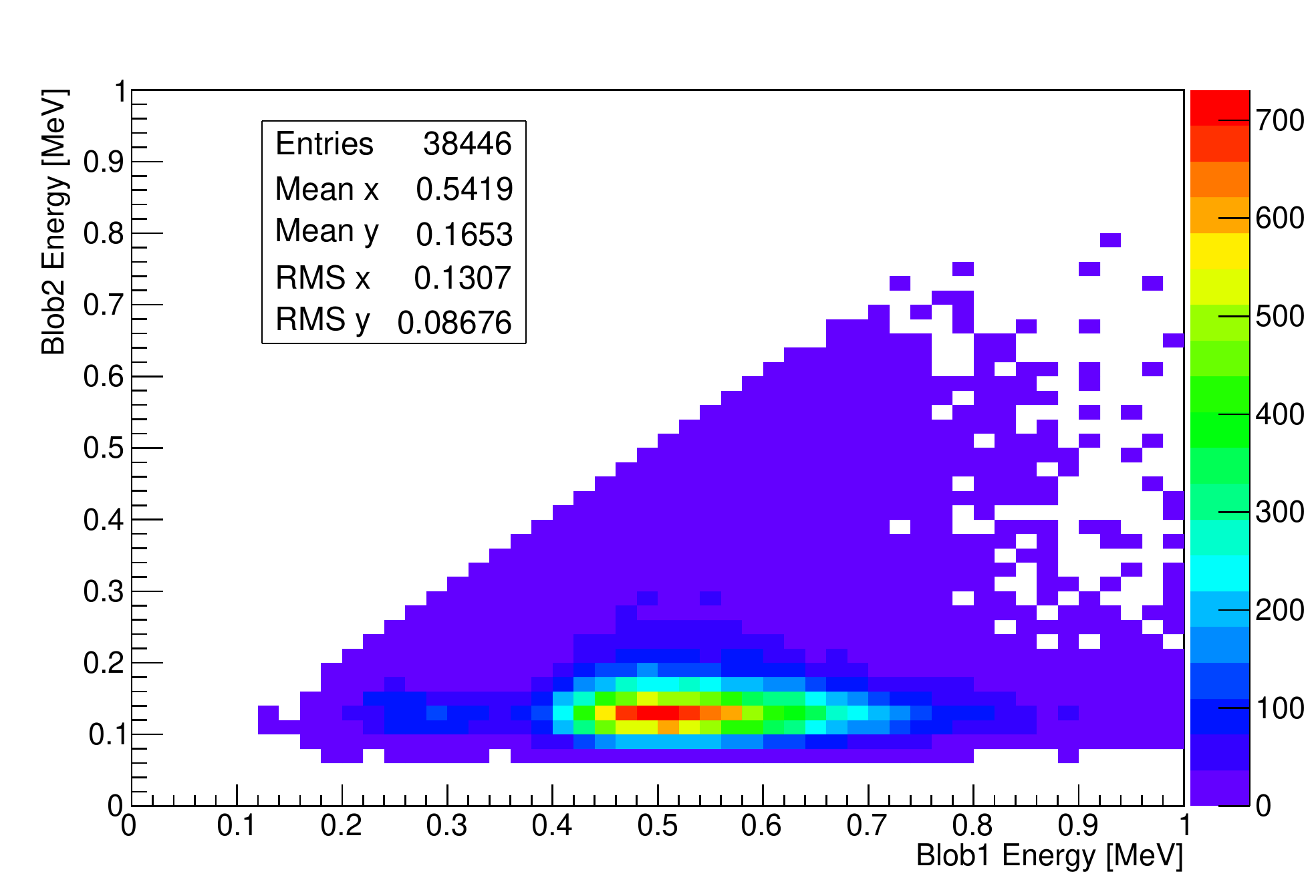}
	\caption{Computed blob candidate energies $E_{b,1}$ vs. $E_{b,2}$ for signal (left) and $^{214}$Bi background (right) events with $10 \times 10 \times 5$ mm$^3$ voxelization (above) and 
		$2 \times 2 \times 2$ mm$^3$ voxelization (below).  The blob radius chosen was $r_b = 18$ mm for the $10 \times 10 \times 5$ mm$^3$~ voxelization and $r_b = 15$ mm for the $2 \times 2 \times 2$ mm$^3$~ voxelization.} \label{fig.blobcuts}
\end{figure}

The results depend on the size of the voxels, which in turn is chosen to reflect the expected performance of the detector under specific operating conditions, as discussed above. Operation with pure xenon corresponds to $10 \times 10 \times 5 \mathrm{~mm^3}$~voxels (conservative), and operation with low diffusion mixtures corresponds to voxel sizes of $2 \times 2 \times 2 \mathrm{~mm^3}$~(best expected case).  Examples of events voxelized with sizes of $10 \times 10 \times 5 \mathrm{~mm^3}$~ and
$2 \times 2 \times 2 \mathrm{~mm^3}$ are shown in Figs. \ref{fig.exampleProjs10105} and \ref{fig.exampleProjs222}.  The histogram of $E_{b,1}$ vs. $E_{b,2}$ is shown in figure 
\ref{fig.blobcuts} for both signal and background events analyzed with both chosen voxel sizes.

Finally we apply a cut designed to choose signal events with two blobs and eliminate background events with only one blob, mandating that $E_{b,1}$ and $E_{b,2}$ are both greater than a threshold energy $E_{\mathrm{th}}$.  This cut is applied to the events remaining after the cut requiring 1 single connected, voxelized track.  For the $10 \times 10 \times 5$ mm$^3$ voxel size with $r_b = 18$ mm and $E_{\mathrm{th}} =$ 0.35 MeV, we eliminate all but 13.3\% of remaining $^{208}$Tl background events and all but 11.0\% of remaining $^{214}$Bi background events, and keep 76.6\% of remaining signal events.  For the $2 \times 2 \times 2$ mm$^3$ voxel size with $r_b = 15$ mm and $E_{\mathrm{th}} =$ 0.3 MeV we eliminate all but 9.74\% of remaining $^{208}$Tl background events and all but 7.55\% of remaining $^{214}$Bi background events, and keep 86.2\% of remaining signal events.  $r_{b}$ was chosen in each case by examining the blob energy with changing $r_b$ and selecting a value large enough to encompass the region of dense energy deposition but small enough to avoid integrating much of the less dense parts of the track.  $E_{th}$ was then varied to give a background rejection near 10\%.

\section{Deep Learning}
The use of artificial neural networks to solve complex problems has been explored since the 1940s.  In recent years, with the dramatic increase in available computing power, the use of computationally
intense neural networks with many inner layers has become feasible.  These neural nets that are many layers deep, called deep neural networks (DNNs), are capable of learning large
amounts of data exhibiting a vast array of features.  This idea of ``deep learning'' has been applied to yield outstanding performance in solving difficult problems such as image \cite{Googlenet} 
and speech \cite{Hinton_2012} recognition.  It has also found recent applications in physics, including event classification in high-energy and neutrino physics experiments \cite{Aurisano_2016, Baldi_2014, deOliveira_2016, Racah_2016}.

Neural networks consist of layers of neurons which compute an output value based on one or several input values.  The output is a function of the weighted sum of the inputs $x_{i}$ plus a bias variable $b$, i.e. $f(\sum_{i}w_{i}x_{i} + b)$, where $f$ is called the activation function and $w_{i}$ are the weights of the neuron, one for each input.  The idea is that
with several layers of many neurons connected together, the values of the final (``output'') layer of neurons will correspond to the solution of some problem given the values input to the
initial layer (called the ``input'' layer).  The weights and biases of all neurons in the network together determine the final output value, and so the network must be trained (the weights and
biases must be adjusted) so that the network solves the correct problem.  This is done by using a training dataset, and for each training event, presenting input data to the network, 
examining its resulting output, and adjusting the weights and biases of the network in a manner that minimizes the discrepancy between the output of the final layer $a$ and the expected 
output $y$.  This adjustment procedure is done by computing a cost function which depends on the actual and expected outputs and quantifies the discrepancy between them, computing 
the gradients of the cost function with respect to the weights and biases in all neurons, and changing the weights and biases in a manner that minimizes the cost function.  After many training
iterations, the weights and biases in the network will ideally have converged to values that not only yield the expected output when the network is presented with an event from the training
dataset, but also yield the expected output when presented with similar events not used in training.  The technical
details behind implementing such a scheme mathematically will not be given here but are discussed at length in \cite{Nielsen_2016}.

\begin{figure}[!htb]
	\centering
	\includegraphics[scale = 0.6]{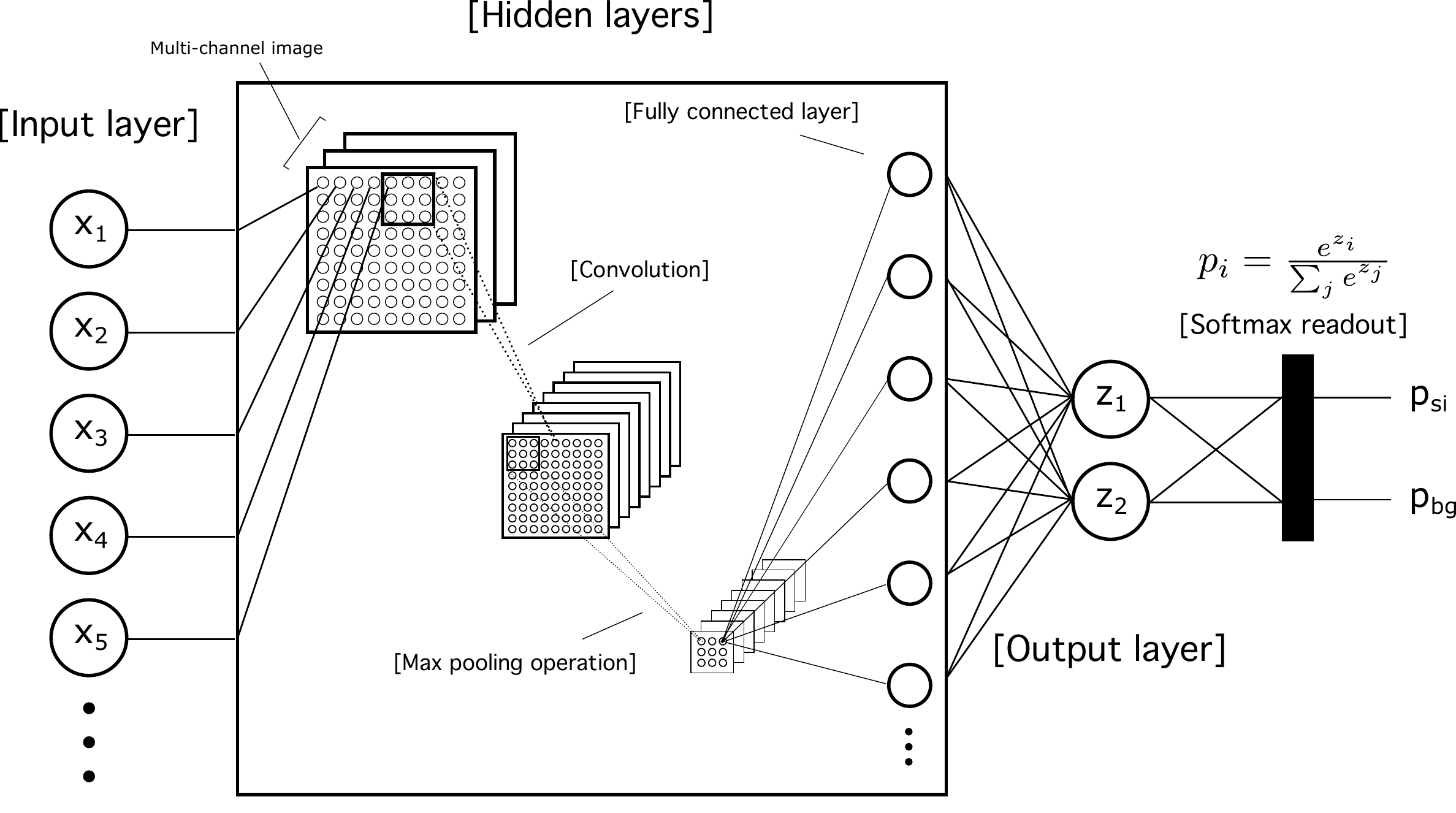}
	\caption{Schematic of a deep convolutional neural network for 2-category classification.  The input layer consists of the pixel intensities of an image, possibly in multiple color channels.  The 
		hidden layers consist of several different operations performed on the input neurons - this example shows a $3 \times 3$ convolution followed by a $3 \times 3$ max-pooling operation, 
		with the resulting neurons input to a fully-connected layer which feeds the two neurons in the output layer.  The activation function of the two neurons in the final layer is such that the two outputs are exponentiated and normalized.  The values in such a layer, called a ``softmax'' readout layer, can then be interpreted as probabilities of classification as
		signal or background.} \label{fig.generalnn}
\end{figure}

Recently, multi-layer convolutional neural networks (CNNs) have been identified as a powerful technique for image recognition problems.  These neural networks consist of
convolutional layers of $n$ columns of $m$ neurons - layers of neurons that share a common set of $m\times n$ weights and a bias.  The set of 
weights $+$ biases is called a filter or kernel, and this filter is combined in a multiplicative sum (a convolution) with an $m\times n$ subset of input neurons to give an output value.  The filter
is moved along the image, each time covering a different $m\times n$ subset of input neurons, and the set of output values corresponding to a single filter is called a feature map.  With this
strategy, further convolutional layers can be used to analyze the higher-level features encoded in the feature maps output
from previous layers.  Often to reduce the amount of computation and neurons present in deeper layers, max-pooling operations are performed, in which the neuron with maximum output
value in an $m\times n$ window (or ``pool'') is selected, and all others in the pool are discarded.  Such an operation performed on a layer of neurons leads to a new layer of reduced size.
A deep CNN may be constructed from a series of several such convolutional operations and max-pooling operations, along with more conventional fully-connected layers, in which all neurons 
output from the previous layer are connected to the input of each neuron in the fully-connected layer, and other operations not discussed here (see figure \ref{fig.generalnn} for a general 
schematic).

In this initial study, we make use of the GoogLeNet \cite{Googlenet}, which is a sophisticated 22-layers-deep convolutional neural network designed for image recognition.  As GoogLeNet was
designed to classify and identify a wide range of features in a full-color images, a more suitable network is likely to exist for our specific problem of classifying particle tracks based on
topology.  While further exploration of DNN architecture is essential to understanding the problem fully, our main goal in this study will be to show that DNNs can ``learn'' to classify 
NEXT events as signal or background potentially better than previously developed conventional analysis methods.

\section{Event classification with a DNN}

Here we investigate the performance of a DNN in classifying events into two categories, ``signal'' and ``background,'' and compare the results to the conventional analysis described in
section \ref{ssec:ClassificationAnalysis}.  We chose to use the GoogLeNet DNN for this initial study, as its implementation was readily available in the Caffe \cite{jia2014caffe}
deep learning framework along with an interface, DIGITS \cite{DIGITS}, which allows for fast creation of image datasets and facilitates their input to several DNN models.  
In order to generate large numbers of events with which to train the DNN, an alternate configuration of the NEXUS Monte Carlo, which we call the ``xenon box'' (Xe box) Monte Carlo, 
was run in which the NEXT-100 detector geometry was not present, and background events (single electrons) and signal events (two electrons emitted from a common vertex with a realistic 
\bbonu\,energy distribution) were generated in a large box of pure xenon gas at 15 bar.  These events were then subject to the same voxelization procedure and single-track cut as described in 
section \ref{ssec.reconstruction}.  

For two different configurations of voxel size, GoogLeNet was trained on 404800 Xe box input events (50\% signal, 50\% background) using one or more NVidia GeForce GPUs.  Each event was input to the net as a .png image 
consisting of three color (RGB) channels, one for each of three projections of the 3D voxelized track, (R, G, B) $\rightarrow$ (xy, yz, xz).  This information for a signal event and a background event
was shown earlier for different voxelizations in Fig. \ref{fig.exampleProjs10105} and Fig. \ref{fig.exampleProjs222}.

\subsection{Analysis of NEXT-100 Monte Carlo}\label{ssec:NEXTMCanalysis}
To compare the ability of the DNN to classify events directly with the performance of the classification analysis of section \ref{ssec:ClassificationAnalysis}, we consider NEXT-100 Monte Carlo 
events that have passed the pre-selection cuts described in \ref{ssec.prep}, with chosen voxel sizes of both $2 \times 2 \times 2$ mm$^3$ and $10 \times 10 \times 5$ mm$^3$.  For each chosen voxel size,
Monte Carlo events that were analyzed with the standard ``blob cuts'' of the classical analysis were classified by the corresponding DNN trained using Xe box events.  Note that the background events
used in this comparison were those produced by $^{214}$ Bi decay generated in the field cage surrounding the active region.   The results are shown in table 
\ref{tbl.DNNcomparison}.  The DNN analysis performs better than the conventional analysis, but there is still potential room for improvement.

\begin{table}[!htb]
	\begin{center}
		\caption[DNN analysis summary]{\label{tbl.DNNcomparison}Comparison of conventional and DNN-based analyses.  The comparison shows, for a given percentage of signal events
			correctly classified, the number of background ($^{214}$Bi) events accepted (mistakenly classified as signal).}
		\begin{tabular}{rcc}
			\\
			\textbf{Analysis} & \textbf{Signal eff. (\%)} & \textbf{B.G. accepted (\%)}\\
			\hline
			DNN analysis ($2 \times 2 \times 2$ voxels) & 86.2 & 4.7\\
			Conventional analysis ($2 \times 2 \times 2$ voxels) & 86.2 & 7.6\\
			\hline
			DNN analysis ($10 \times 10 \times 5$ voxels) & 76.6 & 9.4 \\
			Conventional analysis ($10 \times 10 \times 5$ voxels) & 76.6 & 11.0 \\
		\end{tabular}
	\end{center}
\end{table}

\begin{figure}[!htb]
	\centering
	\includegraphics[scale=0.6]{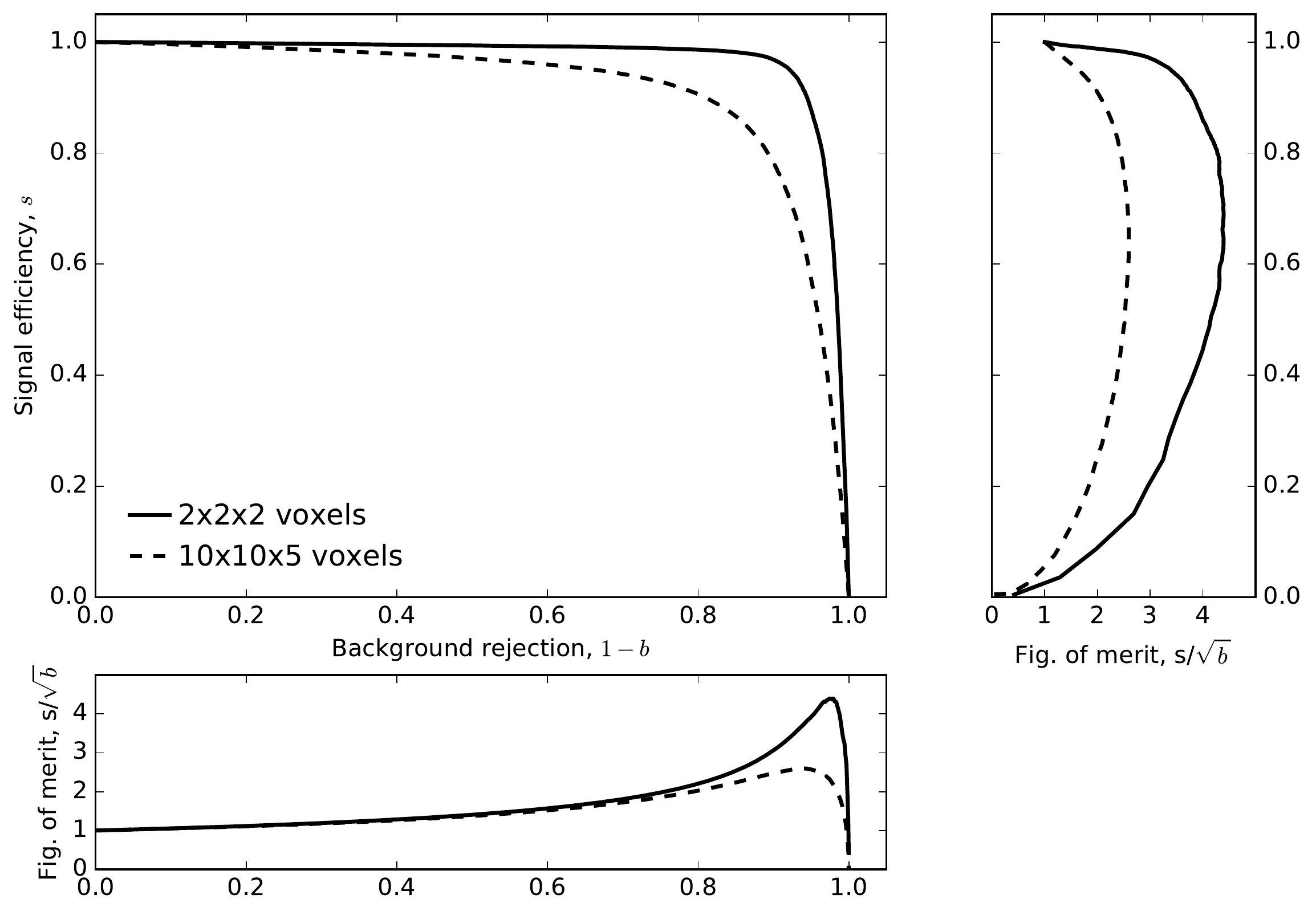}
	\caption{\label{fig_svsb}Signal efficiency vs. background rejection for DNN analysis of voxelized ($2 \times 2 \times 2$ and $10 \times 10 \times 5$ cubic mm), single-track NEXT-100 Monte Carlo events.  The figure of
		merit $F$ to be maximized in an optimal $0\nu\beta\beta$ decay search is also shown as a function of background rejection.}
\end{figure}

Because the output layer of the DNN gives a probability that a given event is signal and a probability that it is background, and these probabilities add to 1, a threshold may be 
chosen for determining whether an event is classified as signal or background.  It can be simply chosen as 50\%, meaning the category with greatest probability is the classification of the
event, or it can be varied to reject further background at the expense of signal efficiency.  Figure \ref{fig_svsb} shows the corresponding pairs of signal efficiency and background 
rejection produced by variation of this threshold, while for the values reported in table \ref{tbl.DNNcomparison} the threshold was chosen such that the signal efficiency matched that reported in 
the conventional analysis.  Note that to optimize the sensitivity to $0\nu\beta\beta$ decay in the case of a non-negligible number ($\gg 1$) of background events, one would seek to maximize 
the ratio of signal events detected divided by the square root of background events accepted (see \cite{NEXT_sensitivity}).  Thus we define a figure of merit $F = s/\sqrt{b}$, where $s$ and 
$b$ are the fractions of signal and background events accepted.  
This quantity is shown alongside the plot of signal efficiency vs. background rejection in Fig. \ref{fig_svsb}.  In table \ref{tbl.DNNcomparison} we reported the values of background rejection 
corresponding to the signal efficiencies studied in the classical analysis, though these did not optimize the figure of merit.  For optimal figures of merit, we would have signal efficiency of 
69.0\% (66.7\%) and background acceptance of 2.5\% (6.6\%) for $2 \times 2 \times 2$ mm$^3$ ($10 \times 10 \times 5$ mm$^3$) voxels.

\begin{figure}[!htb]
	\centering
	\includegraphics[scale=0.55]{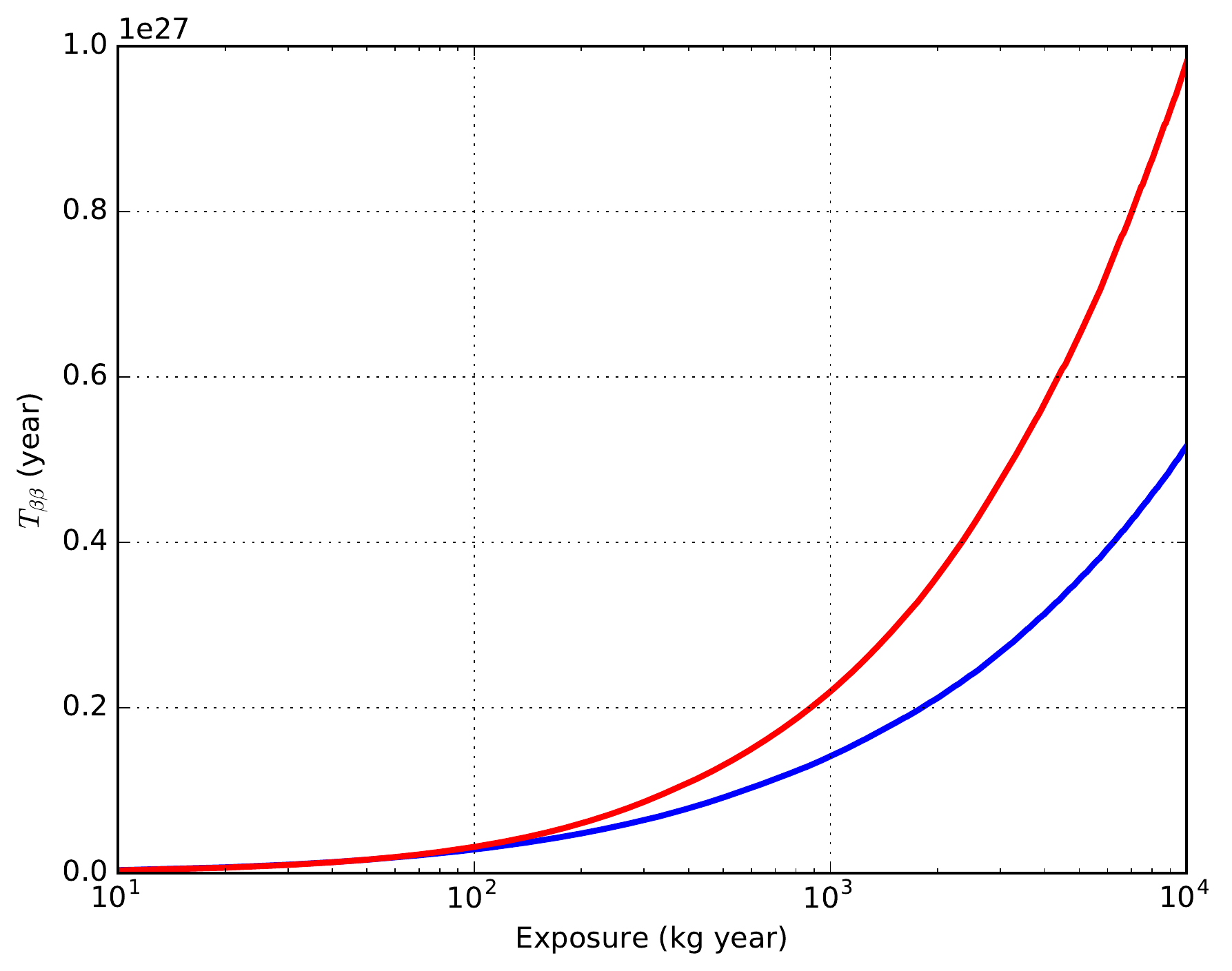}
	\caption{\label{fig.halflife}Sensitivity to the half-life $T^{0\nu}_{1/2}$ calculated using the Feldman-Cousins approach.  The curves describe the NEXT-100 conventional analysis \cite{NEXT_sensitivity} (blue) and NEXT-100 with the improved DNN-based analysis with optimal figure of merit and low diffusion (red).}
\end{figure}

The improvements realized in using the DNN-based analysis combined with lower diffusion translate to significant gains in half-life sensitivity.  Figure \ref{fig.halflife} shows the sensitivity at 90\% confidence level calculated using the Feldman-Cousins \cite{Feldman_1998} prescription as in \cite{NEXT_sensitivity} for the NEXT-100 conventional analysis and for NEXT-100 in the case of low-diffusion ($2 \times 2 \times 2$ mm$^3$ voxels) and using the DNN-based classification with optimal figure of merit. The substantial improvements realizable show the advantages of both an improved DNN-based analysis and achieving low diffusion in NEXT.

\subsection{Evaluating the DNN analysis}\label{ssec:DNNeval}
We now ask what is causing some significant fraction of the events to be misclassified in the analysis described in section \ref{ssec:NEXTMCanalysis}.  To address this, a similar analysis was run 
on several different Monte Carlo datasets generated with differing physics effects, with the goal of developing a better understanding of where potential improvements could be made.

A simple Monte Carlo, which we call the ``toy Monte Carlo'' or ``toy MC,'' was designed to produce ionization tracks of single-electron and two-electron events with a fixed energy
considering minimal physical effects.  Discrete energy depositions were produced with a step size less than 1 mm according to the average stopping power $dE/dx$ as tabulated by
NIST \cite{NIST_mac} for xenon at 15 atm.  Electron multiple scattering was modeled by casting random Gaussian numbers to determine the angles $\theta_{x}$ and $\theta_{y}$ of deflection
from the direction of travel.  Assuming the particle's direction of travel is $\hat{\mathbf{z}}$, the angles $\theta_{x}$ and $\theta_{y}$ between the scattered direction and $\hat{\mathbf{z}}$ projected on 
the x-z and y-z planes respectively, were chosen randomly from a Gaussian distribution with sigma determined according to

\begin{equation}\label{eqn_mscat}
\sigma^{2}(\theta_{x,y}) = \frac{13.6\,\,\mathrm{MeV}}{\beta p}\sqrt{dz/L_{0}}\bigl[1 + 0.038\ln(dz/L_{0})\bigr].
\end{equation}

\noindent where $dz$ is the thickness of xenon travelled in this step, $L_{0}$ is the radiation length in xenon, $p$ is the electron momentum in MeV/c, and $\beta = v/c$, assuming $c = 1$.

Such tracks were generated and voxelized similar to the procedure described in section \ref{ssec.reconstruction}.  Note that no ``single-track'' cut was necessary because no
physics generating a secondary track was implemented.  Also no energy smearing was performed.  For background events, the track generation began with a single electron emitted in a
random direction with energy 2.4578 MeV, while for signal events, this energy was shared equally between two electrons emitted in random directions from a single initial vertex.  The DNN
classified the resulting events with nearly 100\% accuracy, that is, the higher of the two probabilities (> 50\%) of signal or background computed by the DNN corresponded to the correct 
classification in nearly all cases.  Several modifications were then made to attempt to gain insight into the physics causing the lower classification observed in the
more detailed Monte Carlo tracks.  First, a realistic distribution of energies of the two electrons in signal events \cite{Ponkratenko_2000} was used, and later the magnitude of the multiple scattering was doubled (the prefactor 13.6 in equation \ref{eqn_mscat} was increased to 27.2).  The electron energy distribution caused a loss of about 1\% in average accuracy, and the
increased multiple scattering an additional 1\%.  However, even the two effects together were not enough to account for the inaccuracy of about 8\% observed in the events produced by the
full Monte Carlo.

Under the controlled conditions of the Xe box simulation, many events could be generated with different aspects of the physics switched on/off.  It was confirmed that with the same physics as 
that used in the NEXT-100 Monte Carlo, the DNN classified events 
with similar accuracy as before.  Disabling bremsstrahlung seemed to have little effect on the accuracy.  Disabling fluctuations of continuous energy losses in Geant4 had some small 
effect (approx. 1\% increase in accuracy), and disabling the production of secondaries (disallowing the production of secondaries with a range of less than 20 cm) had a more 
significant effect (approx. 2.5\% increase in accuracy), though still did not yield accuracy similar to that of the toy MC datasets.  It was found that disabling both continuous energy fluctuations 
and the production of secondaries gave accuracies similar to that of the toy MC events (about 98\%).  A summary of the key Monte Carlo simulations run and the classification accuracies 
obtained is given in table \ref{tbl.DNNsummary}.

There are two possible scenarios which explain why a DNN misclassifies a particular event. In the first one, the DNN is perfectly capable of taking into account all physical information available in an event but some aspects of the physics of its production have caused it to project the image of an event from the incorrect category (signal or background), given the present detector position and energy resolution. This is, for instance, the case when in a $0\nu\beta\beta$ decay event one of the electrons contains very little energy, and thus it physically resembles a single electron. Likewise, a background event in which a large secondary is produced early on in the production of the single-electron track can look like a two-electron double beta event.

However, it may well be that the present DNN is simply not capable of learning enough information to separate the two types of events, although it is physically possible. This is more difficult to understand by introducing physical effects one at a time. Rather, one should study individual events or construct more complex DNNs until no further improvement appears possible.

\begin{table}[!t]
	\begin{center}
		\caption[DNN analysis summary]{\label{tbl.DNNsummary}Summary of DNN analysis for different Monte Carlo datasets.  The accuracy was computed assuming that the classification
			of the DNN corresponded to the category (signal or background) with the higher ($> 50$\%) probability.  In each case, approximately 15000 signal and 15000 background events were
			used in the training procedure, and between 2000-3000 signal and 2000-3000 background events independent of the training set were used to determine the accuracy.}
		\begin{tabular}{rrc}
			\\
			\textbf{$2 \times 2 \times 2$ voxels} & \textbf{Run description} & \textbf{Avg. accuracy} (\%)\\
			\hline
			\multicolumn{2}{r}{Toy MC, ideal} & 99.8\\
			\multicolumn{2}{r}{Toy MC, realistic \bbonu\,distribution} & 98.9\\
			\multicolumn{2}{r}{Xe box Geant4, no secondaries, no E-fluctuations} & 98.3\\
			\multicolumn{2}{r}{Xe box Geant4, no secondaries} & 94.6\\
			\multicolumn{2}{r}{Xe box Geant4, no E-fluctuations} & 93.0\\
			\multicolumn{2}{r}{Xe box, all physics} & 92.1\\
			\multicolumn{2}{r}{NEXT-100 Geant4} & 91.6\\
			\textbf{$10 \times 10 \times 5$ voxels} & & \\
			\hline
			\multicolumn{2}{r}{NEXT-100 Geant4} & 84.5
		\end{tabular}
	\end{center}
\end{table}

\section{Conclusions}
The NEXT topological signature of $0\nu\beta\beta$ decay events can be used to reject a significant number of background events, thus greatly increasing the sensitivity to $0\nu\beta\beta$ decay.  A DNN-based analysis using GoogLeNet with just three projections seems to be capable of outperforming, by a factor of 1.2 to 1.6, depending on the resolution of reconstruction, in signal/background separation, a conventional analysis based on locating energy ``blobs'' at the ends of the tracks produced by energetic electrons.  The production of secondaries coupled with energy fluctuations in energy deposition seems to be the principle cause of accuracy loss in the DNN analysis.  Future studies geared toward developing a DNN targeted on the problem at hand, possibly exploring fully 3D convolutional networks as opposed to using 2D projections, and attempting to extract information on what characteristics of the tracks it is ``learning,'' would lead to a more complete understanding of the possibilities and limitations of a DNN-based analysis.

\acknowledgments

The NEXT Collaboration acknowledges support from the following agencies and institutions:
the European Research Council (ERC) under the Advanced Grant 339787-NEXT;
the Ministerio de Econom\'{i}a y Competitividad of Spain and FEDER under grants CONSOLIDER-Ingenio
2010 CSD2008-0037 (CUP), FIS2014-53371-C04 and the Severo Ochoa Program
SEV-2014-0398; GVA under grant PROMETEO/2016/120. Fermilab is operated by Fermi Research Alliance, LLC under Contract No. DE-AC02-07CH11359 with the United States Department of 
Energy. JR acknowledges support from a Fulbright Junior Research Award.

\bibliography{dnnext}

\providecommand{\href}[2]{#2}\begingroup\raggedright\begin{thebibliography}{10}

\bibitem{Schechter_1982}
J.~Schechter and J.~W.~F. Valle, \emph{{Neutrinoless double-$\beta$ decay in
  SU(2)$\times$U(1) theories}},
  \href{http://dx.doi.org/10.1103/PhysRevD.25.2951}{\emph{Phys. Rev. D} {\bf
  25} (1982) 2951}.

\bibitem{GomezCadenas:2013ue}
J.~G\'{o}mez-Cadenas, J.~Mart\'{i}n-Albo, J.~Mu\~{n}oz Vidal and C.~Pe\~{n}a
  Garay, \emph{{Discovery potential of xenon-based neutrinoless double beta
  decay experiments in light of small angular scale CMB observations}},
  \href{http://dx.doi.org/10.1088/1475-7516/2013/03/043}{\emph{JCAP} {\bf 1303}
  (2013) 043}, [\href{http://arxiv.org/abs/1301.2901}{{\tt 1301.2901}}].

\bibitem{Cadenas_2012}
J.~J. G{\'o}mez-Cadenas, J.~Mart{\'\i}n-Albo, M.~Mezzetto, F.~Monrabal and
  M.~Sorel, \emph{{The search for neutrinoless double beta decay}},
  \href{http://dx.doi.org/10.1393/ncr/i2012-10074-9}{\emph{Riv. Nuovo Cim.}
  {\bf 35} (2012) 29}, [\href{http://arxiv.org/abs/1109.5515}{{\tt
  1109.5515}}].

\bibitem{Avignone_2008}
F.~T. Avignone~III, S.~R. Elliott and J.~Engel, \emph{{Double beta decay,
  Majorana neutrinos, and neutrino mass}},
  \href{http://dx.doi.org/10.1103/RevModPhys.80.481}{\emph{Rev. Mod. Phys.}
  {\bf 80} (2008) 481}, [\href{http://arxiv.org/abs/0708.1033}{{\tt
  0708.1033}}].

\bibitem{KamLANDZen_2016}
{\scshape KamLAND-Zen} collaboration, A.~Gando, Y.~Gando, T.~Hachiya,
  A.~Hayashi, S.~Hayashida, H.~Ikeda et~al., \emph{Search for majorana
  neutrinos near the inverted mass hierarchy region with {KamLAND-Zen}},
  \href{http://dx.doi.org/10.1103/PhysRevLett.117.082503}{\emph{Phys. Rev.
  Lett.} {\bf 117} (2016) 082503}, [\href{http://arxiv.org/abs/1605.02889}{{\tt
  1605.02889}}].

\bibitem{Gomez-Cadenas:2015twa}
J.~G{\'o}mez-Cadenas and J.~Mart{\'\i}n-Albo, \emph{{Phenomenology of
  neutrinoless double beta decay}},
  \href{http://arxiv.org/abs/1502.00581}{{\tt 1502.00581}}.

\bibitem{NEXT_sensitivity}
{\scshape NEXT} collaboration, J.~Mart\'{i}n-Albo et~al., \emph{{Sensitivity of
  NEXT-100 to neutrinoless double beta decay}},
  \href{http://dx.doi.org/10.1007/JHEP05(2016)159}{\emph{Journal of High Energy
  Physics} {\bf 2016} (2016) 159}, [\href{http://arxiv.org/abs/1511.09246}{{\tt
  1511.09246}}].

\bibitem{Alvarez:2012kua}
{\scshape NEXT} collaboration, V.~{\'A}lvarez et~al., \emph{{Near-intrinsic
  energy resolution for 30 to 662 keV gamma rays in a high-pressure xenon
  electroluminescent TPC}},
  \href{http://dx.doi.org/10.1016/j.nima.2012.12.123}{\emph{Nucl.\ Instrum.\
  Meth.\ A} {\bf 708} (2013) 101}, [\href{http://arxiv.org/abs/1211.4474}{{\tt
  1211.4474}}].

\bibitem{Alvarez:2012xda}
{\scshape NEXT} collaboration, V.~{\'A}lvarez et~al., \emph{{Initial results of
  NEXT-DEMO, a large-scale prototype of the NEXT-100 experiment}},
  \href{http://dx.doi.org/10.1088/1748-0221/8/04/P04002}{\emph{JINST} {\bf 8}
  (2013) P04002}, [\href{http://arxiv.org/abs/1211.4838}{{\tt 1211.4838}}].

\bibitem{Alvarez:2013gxa}
{\scshape NEXT} collaboration, V.~{\'A}lvarez et~al., \emph{{Operation and
  first results of the NEXT-DEMO prototype using a silicon photomultiplier
  tracking array}},
  \href{http://dx.doi.org/10.1088/1748-0221/8/09/P09011}{\emph{JINST} {\bf 8}
  (2013) P09011}, [\href{http://arxiv.org/abs/1306.0471}{{\tt 1306.0471}}].

\bibitem{Lorca:2014sra}
{\scshape NEXT} collaboration, D.~Lorca et~al., \emph{{Characterisation of
  NEXT-DEMO using xenon K$_{\alpha}$ X-rays}},
  \href{http://dx.doi.org/10.1088/1748-0221/9/10/P10007}{\emph{JINST} {\bf 9}
  (2014) P10007}, [\href{http://arxiv.org/abs/1407.3966}{{\tt 1407.3966}}].

\bibitem{NEXT_topology}
{\scshape NEXT} collaboration, P.~Ferrario et~al., \emph{{First proof of
  topological signature in the high pressure xenon gas TPC with
  electroluminescence amplification for the NEXT experiment}},
  \href{http://dx.doi.org/10.1007/JHEP01(2016)104}{\emph{JHEP} {\bf 2016}
  (2016) 104}, [\href{http://arxiv.org/abs/1507.05902}{{\tt 1507.05902}}].

\bibitem{Gotthard_1998}
R.~Luescher, J.~Farine, F.~Boehm, J.~Busto, K.~Gabathuler, G.~Gervasio et~al.,
  \emph{Search for $\beta\beta$ decay in 136{X}e: new results from the
  {G}otthard experiment},
  \href{http://dx.doi.org/10.1016/S0370-2693(98)00906-X}{\emph{Physics Letters
  B} {\bf 434} (1998) 407}.

\bibitem{Cormen_2001}
T.~H. Cormen, C.~Stein, R.~L. Rivest and C.~E. Leiserson, \emph{Introduction to
  algorithms, 2nd ed.}
\newblock McGraw-Hill Higher Education, 2001.

\bibitem{Azevedo:2015eok}
C.~D.~R. Azevedo, L.~M.~P. Fernandes, E.~D.~C. Freitas, D.~Gonzalez-Diaz,
  F.~Monrabal, C.~M.~B. Monteiro et~al., \emph{{An homeopathic cure to pure
  Xenon large diffusion}},
  \href{http://dx.doi.org/10.1088/1748-0221/11/02/C02007}{\emph{JINST} {\bf 11}
  (2016) C02007}, [\href{http://arxiv.org/abs/1511.07189}{{\tt 1511.07189}}].

\bibitem{MartinAlbo_thesis}
J.~Mart\'in-Albo, \emph{{The NEXT experiment for neutrinoless double beta decay
  searches}}.
\newblock PhD thesis, University of Valencia, 2015.

\bibitem{GEANT4}
{\scshape Geant4} collaboration, S.~Agostinelli et~al., \emph{{Geant4---a
  simulation toolkit}},
  \href{http://dx.doi.org/10.1016/S0168-9002(03)01368-8}{\emph{Nucl. Instrum.
  Methods A} {\bf 506} (2003) 250}.

\bibitem{Ponkratenko_2000}
O.~A. Ponkratenko, V.~I. Tretyak and Y.~G. Zdesenko, \emph{{Event generator
  DECAY4 for simulation of double-beta processes and decays of radioactive
  nuclei}}, \href{http://dx.doi.org/10.1134/1.855784}{\emph{Phys. Atom. Nuclei}
  {\bf 63} (2000) 1282}, [\href{http://arxiv.org/abs/nucl-ex/0104018}{{\tt
  nucl-ex/0104018}}].

\bibitem{Googlenet}
C.~Szegedy, W.~Liu, Y.~Jia, P.~Sermanet, S.~E. Reed, D.~Anguelov et~al.,
  \emph{Going deeper with convolutions},
  \href{http://arxiv.org/abs/1409.4842}{{\tt 1409.4842}}.

\bibitem{Hinton_2012}
G.~Hinton et~al., \emph{Deep neural networks for acoustic modeling in speech
  recognition}, \href{http://dx.doi.org/10.1109/MSP.2012.2205597}{\emph{IEEE
  Signal Processing Magazine} {\bf 29} (2012) 82}.

\bibitem{Aurisano_2016}
A.~Aurisano, A.~Radovic, D.~Rocco, A.~Himmel, M.~Messier, E.~Niner et~al.,
  \emph{A convolutional neural network neutrino event classifier},
  \href{http://dx.doi.org/10.1088/1748-0221/11/09/P09001}{\emph{Journal of
  Instrumentation} {\bf 11} (2016) P09001},
  [\href{http://arxiv.org/abs/1604.01444}{{\tt 1604.01444}}].

\bibitem{Baldi_2014}
P.~Baldi, P.~Sadowski and D.~Whiteson, \emph{Searching for exotic particles in
  high-energy physics with deep learning},
  \href{http://dx.doi.org/10.1038/ncomms5308}{\emph{Nature Communications} {\bf
  5} (2014) 1}, [\href{http://arxiv.org/abs/1402.4735}{{\tt 1402.4735}}].

\bibitem{deOliveira_2016}
L.~de~Oliveira, M.~Kagan, L.~Mackey, B.~Nachman and A.~Schwartzman,
  \emph{Jet-images --- deep learning edition},
  \href{http://dx.doi.org/10.1007/JHEP07(2016)069}{\emph{Journal of High Energy
  Physics} {\bf 2016} (2016) 69}, [\href{http://arxiv.org/abs/1511.05190}{{\tt
  1511.05190}}].

\bibitem{Racah_2016}
E.~Racah, S.~Ko, P.~Sadowski, W.~Bhimji, C.~Tull, S.-Y. Oh et~al.,
  \emph{Revealing fundamental physics from the daya bay neutrino experiment
  using deep neural networks},  \href{http://arxiv.org/abs/1601.07621}{{\tt
  1601.07621}}.

\bibitem{Nielsen_2016}
M.~A. Nielsen, \emph{\href{http://neuralnetworksanddeeplearning.com}{Neural
  Networks and Deep Learning}}.
\newblock Determination Press, 2016.

\bibitem{jia2014caffe}
Y.~Jia, E.~Shelhamer, J.~Donahue, S.~Karayev, J.~Long, R.~Girshick et~al.,
  \emph{Caffe: Convolutional architecture for fast feature embedding},
  \href{http://arxiv.org/abs/1408.5093}{{\tt 1408.5093}}.

\bibitem{DIGITS}
{NVIDIA Corporation}, \emph{\href{https://developer.nvidia.com/digits}{{NVIDIA
  DIGITS: Interactive Deep Learning GPU Training System}}}.

\bibitem{Feldman_1998}
G.~D. Feldman and R.~D. Cousins, \emph{A unifed approach to the classical
  statistical analysis of small signals},
  \href{http://dx.doi.org/10.1103/PhysRevD.57.3873}{\emph{Phys. Rev. D} {\bf
  57} (1998) 3873}.

\bibitem{NIST_mac}
J.~H. Hubbell and S.~M. Seltzer,
  \emph{\href{https://www.nist.gov/pml/x-ray-mass-attenuation-coefficients}{NIST:
  Tables of X-Ray Mass Attenuation Coefficients and Mass Energy-Absorption
  Coefficients}}.

\end{thebibliography}\endgroup

\end{document}